# Physics-informed neural networks for programmable origami metamaterials with controlled deployment

Sukheon Kang [a], Youngkwon Kim [b], Jinkyu Yang [b,*], and Seunghwa Ryu [a,c,*]

**Affiliations**

[a] Department of Mechanical Engineering, Korea Advanced Institute of Science and Technology, 291 Daehak-ro, Yuseong-gu, Daejeon 34141, Republic of Korea

[b] Department of Mechanical Engineering, Seoul National University, 1 Gwanak-ro, Gwanak-gu, Seoul 08826, Republic of Korea

[c] KAIST InnoCORE PRISM-AI Center, Korea Advanced Institute of Science and Technology, 291 Daehak-ro, Yuseong-gu, Daejeon 34141, Republic of Korea

[*] Corresponding author e-mail: ryush@kaist.ac.kr (S. Ryu)

[*] Corresponding author e-mail: jkyang11@snu.ac.kr (J. Yang)

**Abstract**

Origami-inspired structures provide unprecedented opportunities for creating lightweight, deployable systems with programmable mechanical responses. However, their design remains challenging due to complex nonlinear mechanics, multistability, and the need for precise control of deployment forces. Here, we present a physics-informed neural network (PINN) framework for both forward prediction and inverse design of conical Kresling origami (CKO) without requiring pre-collected training data. By embedding mechanical equilibrium equations directly into the learning process, the model predicts complete energy landscapes with high accuracy while minimizing non-physical artifacts. The inverse design routine specifies both target stable-state heights and separating energy barriers, enabling freeform programming of the entire energy curve. This capability is extended to hierarchical CKO assemblies, where sequential layer-by-layer deployment is achieved through programmed barrier magnitudes. Finite element simulations and experiments on physical prototypes validate the designed deployment sequences and barrier ratios, confirming the robustness of the approach. This work establishes a versatile, data-free route for programming complex mechanical energy landscapes in origami-inspired metamaterials, offering broad potential for deployable aerospace systems, morphing structures, and soft robotic actuators.

1. **Introduction**

Origami-inspired mechanical systems offer a unique path to lightweight yet multifunctional structures, combining high reconfigurability with geometric design freedom [1,2]. This combination enables motion profiles, deployment behaviors, and adaptive functions that conventional engineering designs cannot easily achieve, opening new opportunities across aerospace [3,4], robotics [5-7], energy devices [8-11], and mechanical metamaterials [12,13]. Among the various origami motifs, Kresling origami originates from the periodic wrinkle patterns formed during torsional buckling of thin cylindrical shells and exhibits a distinctive kinematic coupling between axial contraction and twisting during folding [14]. By tuning only a few geometric parameters, Kresling origami can exhibit multistability, with its energy landscape programmable purely through geometry [15-18]. These properties have led to its use in deployable structures, soft robotic actuators, impact-mitigation devices, reconfigurable antennas, and other advanced systems [19-25].

Understanding and exploiting these unique mechanical responses requires accurate modeling approaches and efficient design strategies. To model and design Kresling origami, most previous studies have used truss-based or continuum mechanics approaches combined with conventional optimization algorithms [26-29]. While effective for fixed geometries or targeting specific stable states, these methods face significant challenges in addressing high-dimensional objectives, such as tailoring nonlinear energy landscapes, achieving multistability, or programming precise deployment sequences, and they often lose generality when applied to more complex architectures.

Recent advances in machine learning offer complementary pathways to overcome these modeling limitations. Data-driven approaches have emerged, training deep learning models on large finite element datasets to predict or inversely design the mechanical behavior of architected metamaterials [30-33], including origami structures [34-37]. However, these approaches require extensive labeled datasets and often fail to generalize beyond the predefined design space. Physics-informed neural network (PINN) approaches address this limitation by embedding governing equations directly into the learning process [38], enabling design without labeled data. PINNs have been applied to topology optimization, metamaterial design, and energy-based structural tailoring [39-43].

Within this growing field of research, early efforts have begun to explore PINN-based origami design. For Kresling origami, Liu et al. [44] made an important first step by applying a PINN-based approach to predict the energy curve of cylindrical configurations and demonstrate a basic inverse design. Their framework, designed for axially symmetric geometries, focused on matching discrete points on the energy curve and did not aim to control barrier heights or shape the complete energy landscape required for prescribed deployment paths. Finite element analysis (FEA) and experimental validation were beyond the scope of that study. Furthermore, the concept of assembling multilayer structures and programming energy barriers in a hierarchical manner to achieve sequential deployment remains largely unexplored, despite its importance for deployable antennas, morphing aerospace skins, and continuously actuated soft robotic systems.

Building on these foundations, the present study introduces an integrated PINN framework for both forward prediction and inverse design of conical Kresling origami (CKO). The framework predicts full energy landscapes without pre-collected training data and directly maps geometric parameters to mechanical responses. Enhanced static-equilibrium constraints allow generalization beyond cylindrical to arbitrary conical geometries while preventing non-physical artifacts. The inverse design routine specifies both the target heights of stable states and the magnitudes of the separating energy barriers, enabling freeform control over the entire energy curve. Leveraging this capability, we design hierarchical CKO assemblies with programmed barriers that achieve controlled, layer-by-layer deployment. The approach is validated through FEA and physical prototypes, demonstrating robust agreement in deployment order and barrier ratios.

## 2. Methods

### 2.1. Geometric, kinematic, and energy modeling of conical Kresling origami

The CKO structure consists of periodically arranged quadrilateral unit cells folded from a flat crease pattern (Figure 1(A), (B)). Each unit cell is defined by five geometric parameters: the top and bottom edge lengths $a$ and $b$, the mountain and valley crease lengths $c$ and $d$, and the internal angle $\beta$ between the bottom edge and the adjacent mountain crease. By folding the mountain and valley creases of *n* unit cells and connecting them along the lateral edges, the structure forms a conical shell in which the top and bottom polygonal cross-sections differ in radius [45,46]. This configuration induces a characteristic deformation mode in which axial contraction is coupled with rotational twisting.

The deformation of the structure is fully described by two state variables: the vertical height $h$ between the polygons and the in-plane rotation angle $\varphi$ of the top polygon relative to the bottom. We define $\varphi = 0$ as the reference configuration in which the side triangles are symmetrically aligned without in-plane twist, and the structure exhibits no chirality. As shown in Figure 1(C), compression leads to simultaneous reduction in height and increase in rotation, producing smooth and continuous folding motion. For clarity, both physical paper models and their truss abstractions are shown at four representative deformation stages.

To model the mechanical response, we adopt a truss-based abstraction in which each mountain and valley crease is treated as a linear axial spring resisting extension and compression. We note in passing that the CKO is a non-rigid origami, experiencing evident facet deformation during folding motions. Thus, this axial spring model not only represents the crease contraction/extension, but also encompasses more complicated facet and crease deformation (e.g., bending and twisting motions of facets, and axial, shear and torsional motions of creases). Under this assumption, the total strain energy of the structure is computed as the sum of elastic energies stored in the creases:

$$U(h, \varphi) = \frac{n}{2} k_m (L_m - L_m^0)^2 + \frac{n}{2} k_v (L_v - L_v^0)^2 \qquad (1)$$

Here, $L_m$ and $L_v$ denote the current lengths of the mountain and valley creases, and $L_m^0$, $L_v^0$ are their rest lengths in the unfolded state. The stiffness values $k_m = EA_m/L_m^0$ and $k_v = EA_v/L_v^0$ are defined by the Young's modulus $E$ and the cross-sectional areas $A_m$ and $A_v$ of the axial truss members, which are normalized to unity in this work.

The resulting energy landscape $U(h, \varphi)$ is a two-dimensional surface composed of valleys and ridges that capture the multistable nature of the structure. As shown in Figure 1(D), the red dashed curves trace equilibrium paths where the torque vanishes, i.e., $\partial U/\partial \varphi = 0$. These represent the physically realizable rotation states at each height. Figure 1(E) presents the energy and rotation profiles along the minimum-energy branch, where the system follows the lowest-energy trajectory under axial loading. The four deformation stages in Figure 1(C) are marked for reference. This framework captures essential CKO mechanics and forms the physical basis for the learning models described in the next section. Full derivations are provided in Supplementary Information Section 1.

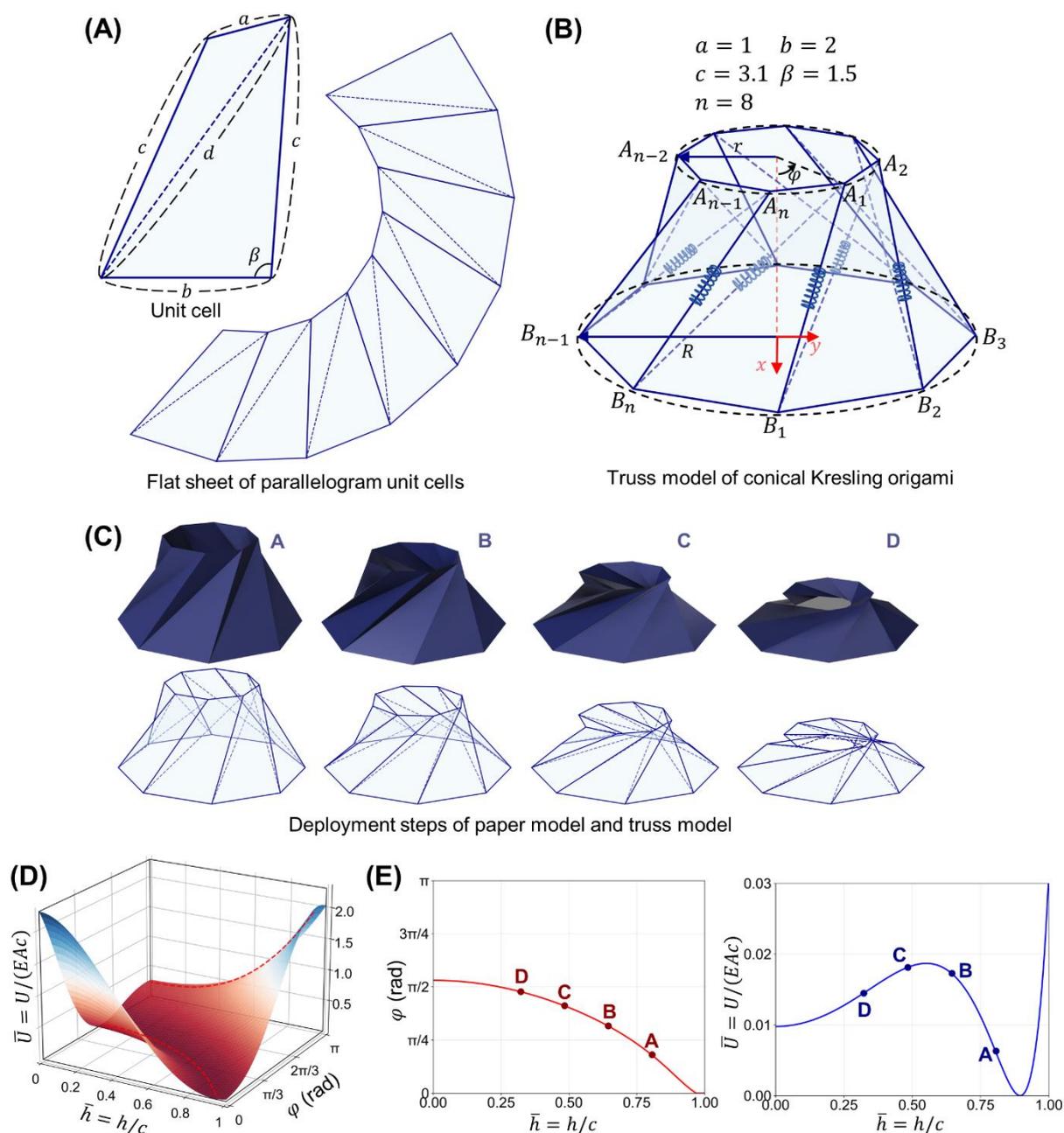

**Figure 1.** Geometric modeling and energy landscape of conical Kresling origami (CKO). (A) Unit cell and flat crease pattern with geometric parameters: $a, b, c, d,$ and $\beta$. (B) 3D CKO structure and truss model showing vertices $(A_i, B_i)$, radii $(r, R)$, and rotation angle $\varphi$ ($a = 1, b = 2, c = 3.1, \beta = 1.5, n = 8$). (C) Compression sequence (A-D) with paper model (top) and truss model (bottom). (D) Energy landscape $U(\bar{h}, \varphi)$ with torque-free equilibrium paths (red dashed lines). (E) Energy $U$ and rotation angle $\varphi$ versus normalized height $\bar{h}$, with points A-D corresponding to stages in (C).

## 2.2. Physics-informed neural network framework

### 2.2.1. Forward prediction using PINN

In the forward design problem, the goal is to predict the potential energy landscape of a CKO with given geometry by learning the deformation-dependent rotation $\varphi(h)$ using a PINN. The neural network architecture, shown in Figure 2(A), is a fully connected feedforward model with two hidden layers of 128 tanh-activated neurons. The input layer receives the height $h$, while geometric parameters $(a, b, c, \beta, n)$ are treated as fixed constants. To ensure physical feasibility, the rotation angle $\varphi$ is constrained to $[0, \varphi_{upper}]$ through a sigmoid-based scaling:

$$\varphi(h) = \varphi_{lower} + (\varphi_{upper} - \varphi_{lower}) \cdot \sigma(z(h)) \quad (2)$$

where $z(h)$ is the raw network output, $\sigma(\cdot)$ is the sigmoid function, and $\varphi_{upper} = \min(\varphi_f, \varphi_s)$ with $\varphi_f$ being the folded-flat angle and $\varphi_s = \pi - 2\pi/n$ the locked configuration.

The network is optimized using a variational objective functional based on minimum total potential energy:

$$J[\varphi] = \int_{h_{min}}^{h_{max}} U(h, \varphi(h)) \, dh \quad (3)$$

With an Euler–Lagrange (EL) regularization term:

$$\frac{\partial U}{\partial \varphi} - 2\lambda_{reg} \frac{d^2\varphi}{dh^2} = 0 \quad (4)$$

The total loss function combines:

$$L_{total} = L_E + \lambda_{EL} L_{EL} \quad (5)$$

$$L_E = \frac{1}{m}\sum_{i=1}^{m} U(h_i, \varphi(h_i)) \tag{6}$$

$$L_{EL} = \frac{1}{m}\sum_{i=1}^{m}\left(\frac{\partial U}{\partial \varphi}(h_i) - 2\lambda_{reg}\frac{d^2\varphi}{dh^2}(h_i)\right)^2 \tag{7}$$

The energy loss $L_E$ minimizes total potential energy, while $L_{EL}$ promotes smoothness in $\varphi(h)$ and prevents nonphysical discontinuities in multistable systems. The detailed training procedure is provided in Algorithm 1 (Supplementary Information Section 2).

### 2.2.2. Inverse design based on target energy curve

In inverse design, both the structural parameters $(a, b, c, \beta)$ and $\varphi(h)$ are optimized to match a target energy profile. The network architecture remains identical to the forward case, but structural parameters are now encoded as differentiable variables constrained via logarithmic and sigmoid transformations (Figure 2(B)).

The total loss function is:

$$L_{total} = \lambda_{target}L_{target} + \lambda_{phys}L_{phys} + \lambda_{EL}L_{EL} + L_{penalty} \tag{8}$$

$$L_{target} = \frac{1}{m}\sum_{i=1}^{m}(U(h_i, \varphi(h_i)) - U_{target}(h_i))^2 \tag{9}$$

$$L_{phys} = \frac{1}{m}\sum_{i=1}^{m}(\frac{\partial U}{\partial \varphi}(h_i))^2 \tag{10}$$

$$L_{penalty} = \begin{cases} 0, & \text{physical constraints are satisfied} \\ C, & \text{otherwise} \end{cases} \quad with\ C \gg 1(e.g., C = 10^3) \tag{11}$$

Here, $L_{target}$ enforces energy curve matching, $L_{phys}$ ensures torque-free equilibrium ($\partial U/\partial \varphi = 0$), and $L_{EL}$ (as defined in Eq. 7) maintains variational consistency. The penalty term $L_{penalty}$ enforces geometric feasibility:

$$|r - R| \leq c \leq r + R \text{ (Mountain crease feasibility)}$$
$$d > 0 \text{ (Valley crease positivity)}$$
$$\beta \in (0, \pi) \text{ (Angle validity)} \quad (12)$$
$$-1 \leq \cos(\varphi_f) \leq 1 \text{ (Flat foldability condition)}$$

The training procedure (Algorithm 2, Supplementary Information Section 2) iteratively updates both parameters and network weights via gradient descent.

### 2.2.3. Inverse design based on bistable energy programming

This framework extends inverse design to underdetermined problems requiring only minimal specifications: two stable heights $(h_1, h_2)$ with zero energy and a target barrier value $U_{barrier}$. The barrier location is autonomously inferred by the model.

The total loss function incorporates additional terms:

$$L_{total} = \lambda_{target}L_{target} + \lambda_{phys}L_{phys} + \lambda_{EL}L_{EL} + \lambda_{barrier}L_{barrier}$$
$$+ \lambda_{smooth}L_{smooth} + L_{penalty} \quad (13)$$

$$L_{target} = \frac{1}{k}\sum_{i=1}^{k}(U(h_i, \varphi(h_i)) - U_{target}(h_i))^2 \quad (14)$$

where $L_{phys}$ (Eq. 10) and $L_{EL}$ (Eq. 7) remain as previously defined. The novel barrier constraint loss ensures proper extremum identification:

$$L_{barrier} = L_{barrier,1} + L_{barrier,2} \quad (15)$$

$$L_{barrier,1} = \left|\frac{\partial U}{\partial h}\right|_{h=h_{max}}^{2} \tag{16}$$

$$L_{barrier,2} = \text{ReLU}\left(\frac{\partial^2 U}{\partial h^2}\bigg|_{h=h_{max}}\right) \tag{17}$$

where $h_{max} = argmax\{U(h)|h \in [h_1, h_2]\}$.

$$L_{smooth} = \frac{1}{m-2}\sum_{i=1}^{m-2}\left(\frac{\partial^2 U}{\partial \varphi^2}(h_i)\right)^2 \tag{18}$$

$L_{barrier,1}$ enforces zero gradient at the barrier, while $L_{barrier,2}$ ensures convexity via ReLU activation. $L_{smooth}$ globally regularizes energy curve curvature. The constraint penalty follows Eq. 11 with conditions from Eq. 12. Figure 2(B) illustrates the complete architecture, with training details in Algorithm 3 (Supplementary Information Section 2).

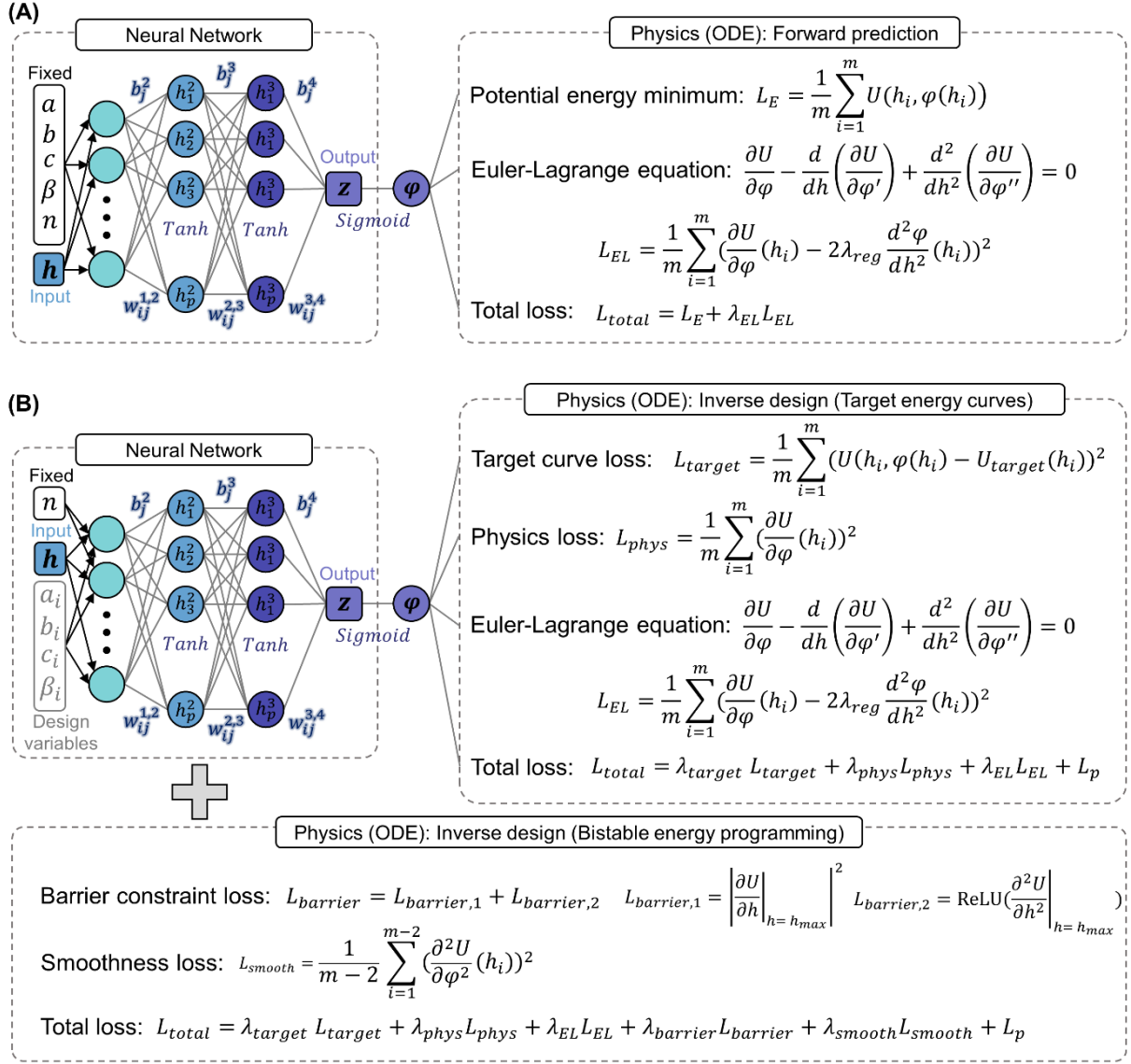

**Figure 2.** Physics-informed neural network (PINN) framework. (A) Forward prediction: Network predicts $\varphi(h)$ from height h with energy minimization ($L_E$) and Euler-Lagrange regularization ($L_{EL}$). (B) Inverse design: Joint optimization of structural parameters ($a, b, c, \beta$) and $\varphi(h)$. Base loss includes $L_{target}$, $L_{phys}$, $L_{EL}$, and $L_{penalty}$. For bistable programming, additional $L_{barrier}$ and $L_{smooth}$ terms enable minimal specification design.

### 2.3. Experimental methods

Three zigzag type CKO prototypes (layer 1, 2, and 3) were fabricated using 0.2 mm-thick polyethylene terephthalate (PET) sheet. The top and bottom polygonal plates were made of 2 mm-thick acrylic plates. All PET and acrylic were cut by a laser cutting machine (Trotec Speedy 360). PET sheets were connected to the acrylic plates with M3 bolts and nuts, and the two ends of each PET panel joints were joined using double-sided foam tape.

The geometric configuration of the prototypes was a zigzag CKO with a fixed polygonal order of $n = 8$ (regular octagon). Edge lengths $a = 24.5$ mm and $b = 49$ mm were constant, while the mountain crease length $c$ and the internal angle $\beta$ varied as shown in Figure 3(A): layer 1 ($c = 83.1$ mm, $\beta = 87.6°$), layer 2 ($c = 85.9$ mm, $\beta = 85.5°$), layer 3 ($c = 87.8$ mm, $\beta = 83.8°$).

To improve foldability and durability, a double-stitch perforation (DSP) pattern was applied based on the design methodology proposed in previous studies [47]. The DSP pattern permits shear and twist along the crease, reduces fatigue and fracture, and is well- suited for non-rigid origami mechanisms such as the CKO. The DSP parameters ($L, B, H, W$) were based on the bi-stable configuration of Kresling structures. Parameters ($L, B, H, W$) = (10, 1.0, 0.4, 0.6) mm for the $a$ and $b$ creases, and (10, 1.0, 0.4, 0.4) mm were for the $c$ and $d$ creases as shown in Figure 3(A).

Experimental tests were conducted to validate the theoretical model and examine the mechanical behavior of the CKO. The experimental setup, shown in Figure 3(B), consisted of a CKO mounted on an in-house load frame with displacement controlled by a linear stage (Velmex Inc., Bislide MN10). Force was measured using a tension/compression load cell (LUX-B-200N-ID), with a sleeve bearing placed below the load cell to allow free rotation of the top plate and enable the natural twisting motion of the CKO.

Two specimen types were tested: a single-layer CKO and a multi-layer CKO assemblies. For each type, uniaxial compression tests were performed three times, and the averaged force data were used for analysis. Force–displacement curves were obtained directly from the measured data, and energy–displacement curves were calculated by numerical integration of the force over displacement.

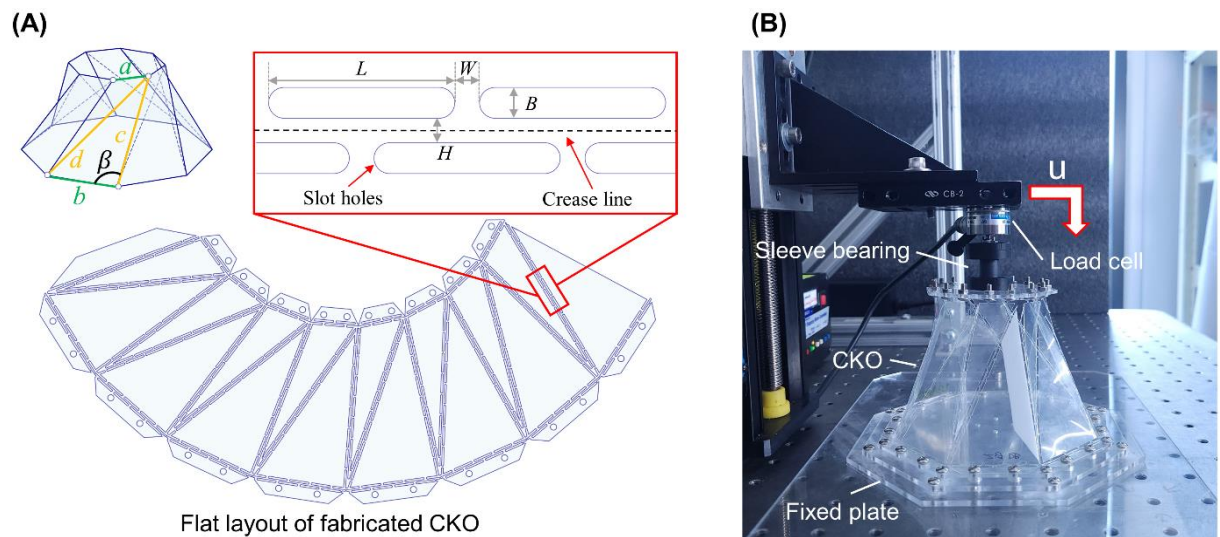

**Figure 3.** Geometry and experimental setup of a CKO. (A) Geometric design and crease parameters of the CKO unit. (B) Experimental setup for uniaxial compression test of the CKO.

### 2.4. Finite Element Analysis

To validate the predicted energy responses of CKO structures, FEA was conducted using the ABAQUS/Implicit solver from Dassault Systèmes. The simulation employed a beam-truss model consistent with the truss-based formulation described in Section 2.1. Mountain and valley creases were modeled as linear elastic beam elements with Young's modulus $E = 1 MPa$, cross-sectional area $A = 20\ mm^2$, and Poisson's ratio 0.3 [45]. To simulate rigid behavior, the top and bottom polygonal edges were assigned a modulus 100 times greater than that of the crease elements. The geometric dimensions were uniformly scaled by a factor of 100 relative to the nominal dimensions used in the PINN framework.

Two types of simulations were performed depending on the configuration. In the single-layer CKO configuration, energy-versus-height profiles $U(h)$ were obtained by first compressing the structure from its zero-energy equilibrium height to a fully folded state, and then gradually stretching it while recording the energy over a range of increasing heights. During this process, the bottom polygon was fixed, and displacement control was applied to the top polygon.

For multi-layer CKO assemblies, the focus was placed on measuring energy as a function of height change rather than absolute height. Each structure was compressed from its equilibrium state, and the total strain energy was computed over the deformation process. This approach captures the energy evolution associated with sequential deployment under external loading.

The FEA results were quantitatively compared with those obtained from the PINN-based models introduced in Sections 2.2.1 through 2.2.3, including forward predictions, inverse designs, and energy barrier programming in multi-unit assemblies.

## 3. Results

### 3.1. Forward prediction of energy Curve

We first evaluate the forward prediction capability of the proposed PINN framework by testing its accuracy in predicting the deformation-dependent energy response $\bar{U}(\bar{h})$ for CKO structures with diverse geometries. Figure 4(A) shows the energy landscape computed from the truss-based model over the normalized design space ($\bar{h}, \varphi$), with torque-free equilibrium paths ($\partial U/\partial \varphi = 0$). The PINN-predicted paths (blue dashed) coincide with these equilibrium curves across all cases, confirming that the model identifies physically valid solutions solely from embedded mechanical constraints without labeled data.

To examine the predicted energy curves in more detail, Figure 4(B) compares the PINN predictions (blue solid lines) with FEA results (black dots) for four representative geometries chosen to illustrate distinct energy landscape features.

- Case1: classical bistability with two zero-energy minima.
- Case2: near-monostability with a single dominant zero-energy minimum.
- Case3: asymmetric bistability, with one zero-energy minimum and another at a non-zero energy level.
- Case4: presents a zero-energy stable state near $\bar{h} = 0$ (fully folded configuration) and another at higher $\bar{h}$, serving as a critical structure for the inverse design and sequential deployment demonstrations

In Figure 4B, red markers identify the zero-energy stable points, with corresponding 3D renderings shown in red. Green markers indicate local maxima (energy barriers) whose 3D configurations are shown in green. These annotations highlight the diversity and programmability of the CKO energy landscape, from symmetric bistability to highly asymmetric responses.

In all cases, the predicted curves closely follow the FEA results, capturing both the shape of the energy profile and the positions of stable and unstable states. Minor discrepancies are negligible and do not affect the qualitative features of the energy landscape. Quantitatively, Table 1 reports the root mean square error (RMSE) and coefficient of determination ($R^2$) between PINN and FEA results, with all four cases achieving $R^2 > 0.93$ and RMSE values within $5 \times 10^{-3}$. These results confirm that the proposed forward PINN framework can accurately reproduce a wide variety of stability behaviors, including both bistable and monostable configurations, as dictated by the underlying geometry.

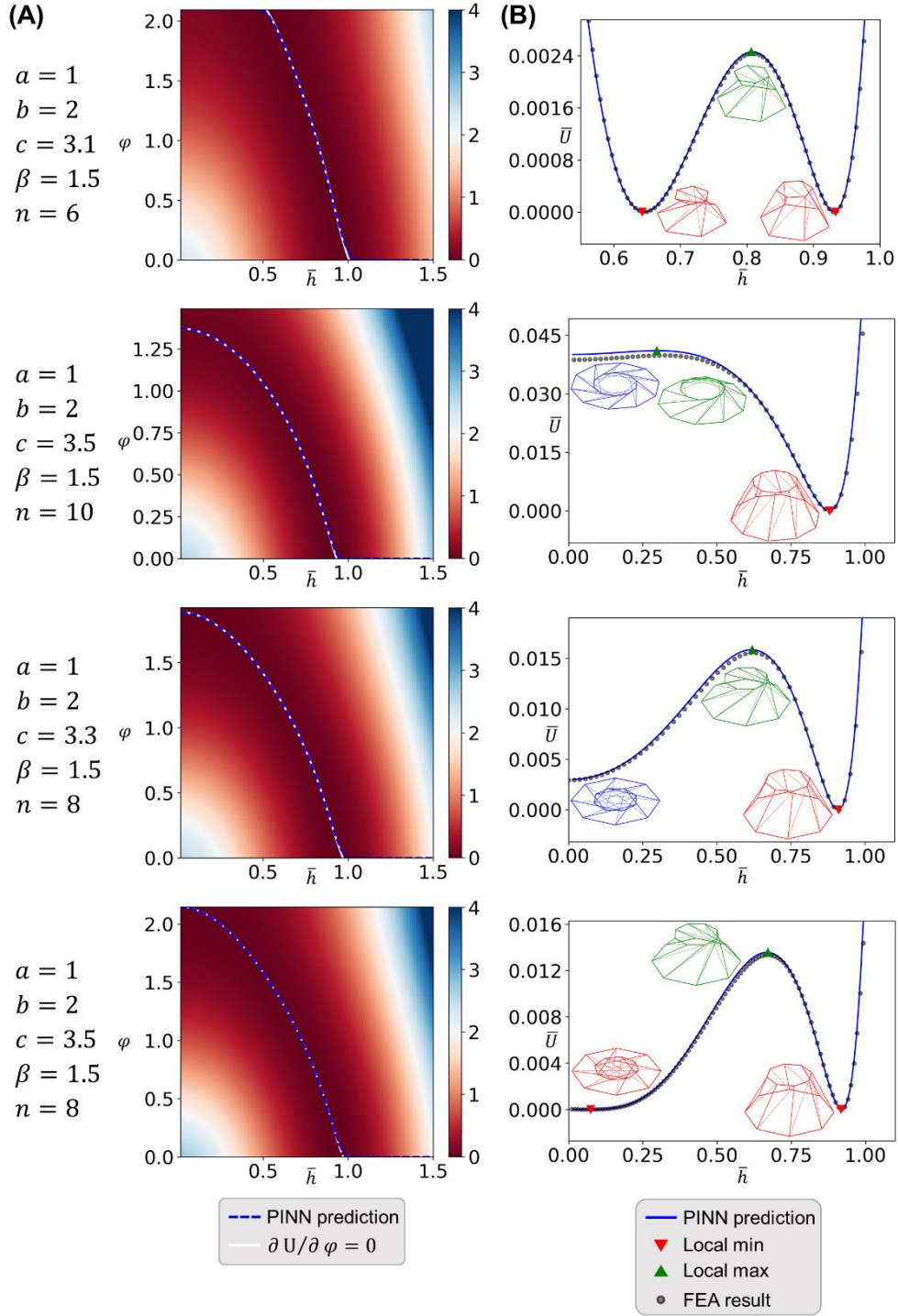

**Figure 4.** Forward prediction of energy landscapes for diverse CKO geometries. (A) Energy fields $\overline{U}(\overline{h}, \varphi)$ for four representative cases with torque-free equilibrium paths (white solid lines: $\partial U/\partial \varphi = 0$) and PINN predictions (blue dashed lines). (B) Energy curves $\overline{U}(\overline{h})$ showing PINN predictions (blue solid lines) versus FEA results (black dots). Red triangles mark stable minima and green triangles indicate energy barriers, with corresponding 3D CKO structures embedded at these critical points. From top to bottom: Case 1 exhibits bistability with two zero-energy states; Case 2 shows monostability; Case 3 demonstrates asymmetric bistability with non-zero energy minimum (blue structure); Case 4 shows bistability with zero-energy states, one at $\overline{h} = 0$ (fully folded).

| Case | $R^2$ (FEA) | RMSE (FEA) |
|---|---|---|
| Case 1 | 0.991081 | 0.000216 |
| Case 2 | 0.930294 | 0.004485 |
| Case 3 | 0.997174 | 0.000256 |
| Case 4 | 0.973334 | 0.001053 |

**Table 1.** Forward prediction accuracy for four representative CKO geometries. Coefficient of determination ($R^2$) and root mean square error (RMSE) values compare PINN predictions with FEA results for the normalized energy curves shown in Figure 4(B).

### 3.2. Inverse design based on full energy curve

We then evaluate the inverse design capability of the proposed PINN framework by reconstructing CKO geometries that reproduce a prescribed target energy curve $\overline{U}(\overline{h})$. In contrast to the forward problem, where geometry is fixed and the deformation field is learned, the inverse design task treats both the structural parameters and the deformation field as trainable variables optimized simultaneously.

Figure 5(A) presents the results for the four representative cases introduced in the forward prediction study. In each case, the target curve (red solid line), the PINN-predicted curve for the optimized geometry (blue dashed line), and the FEA validation results (black dots) are plotted over the normalized height domain. Across all cases, the three curves are in excellent agreement, indicating that the PINN-based inverse framework reliably identifies geometries whose mechanical response matches the desired target profile.

The optimized structural parameters for the inverse-designed geometries differ from those used to generate the target curves, reflecting the one-to-many mapping between energy landscape and geometry. Despite these differences in parameter values (Supplementary Table S1), the resulting energy curves align almost perfectly with the targets, and FEA verification confirms that the designed structures are physically valid. Quantitatively, all cases achieve $R^2 > 0.99$ with respect to the target curves and $R^2 > 0.99$ in FEA validation, with RMSE values well below $1 \times 10^{-3}$ (Supplementary Table S2).

Figure 5(B) shows the convergence behavior of the loss functions during training. For each of the four cases, the curves for the total loss, Euler–Lagrange loss, target matching loss, and physics constraint loss are plotted. The loss values decrease rapidly in the early stages and then enter a gradual refinement phase, eventually converging stably. In addition, Figure 6 presents the evolution of the structural parameters $(a, b, c, \beta)$ over the course of training. All four parameters exhibit stable convergence as training progresses, demonstrating the numerical stability and reliability of the proposed inverse design framework.

Overall, these results confirm that the proposed PINN-based inverse design framework can robustly reconstruct physically valid CKO geometries that reproduce prescribed energy curves with high accuracy, even when the optimal geometry differs from the original configuration.

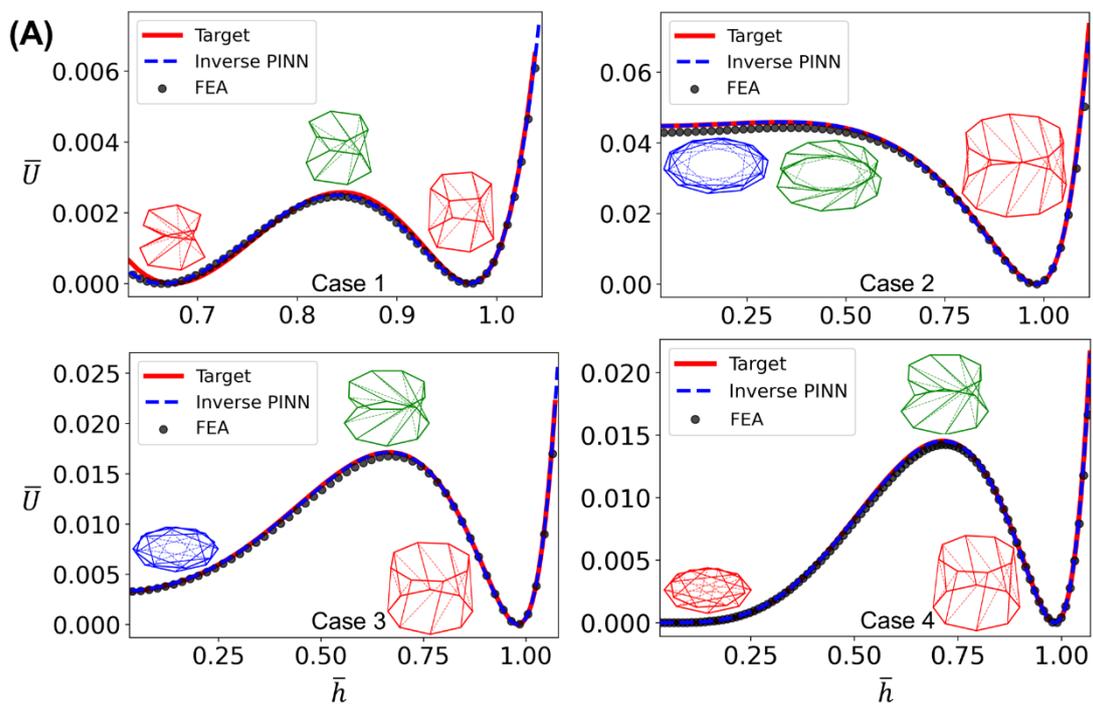

**Figure 5.** Inverse design based on target energy curves. Comparison of normalized energy curves $\bar{U}(\bar{h})$ for four representative cases, showing target profiles (red solid lines), inverse-designed PINN predictions (blue dashed lines), and FEA validation results (black dots).

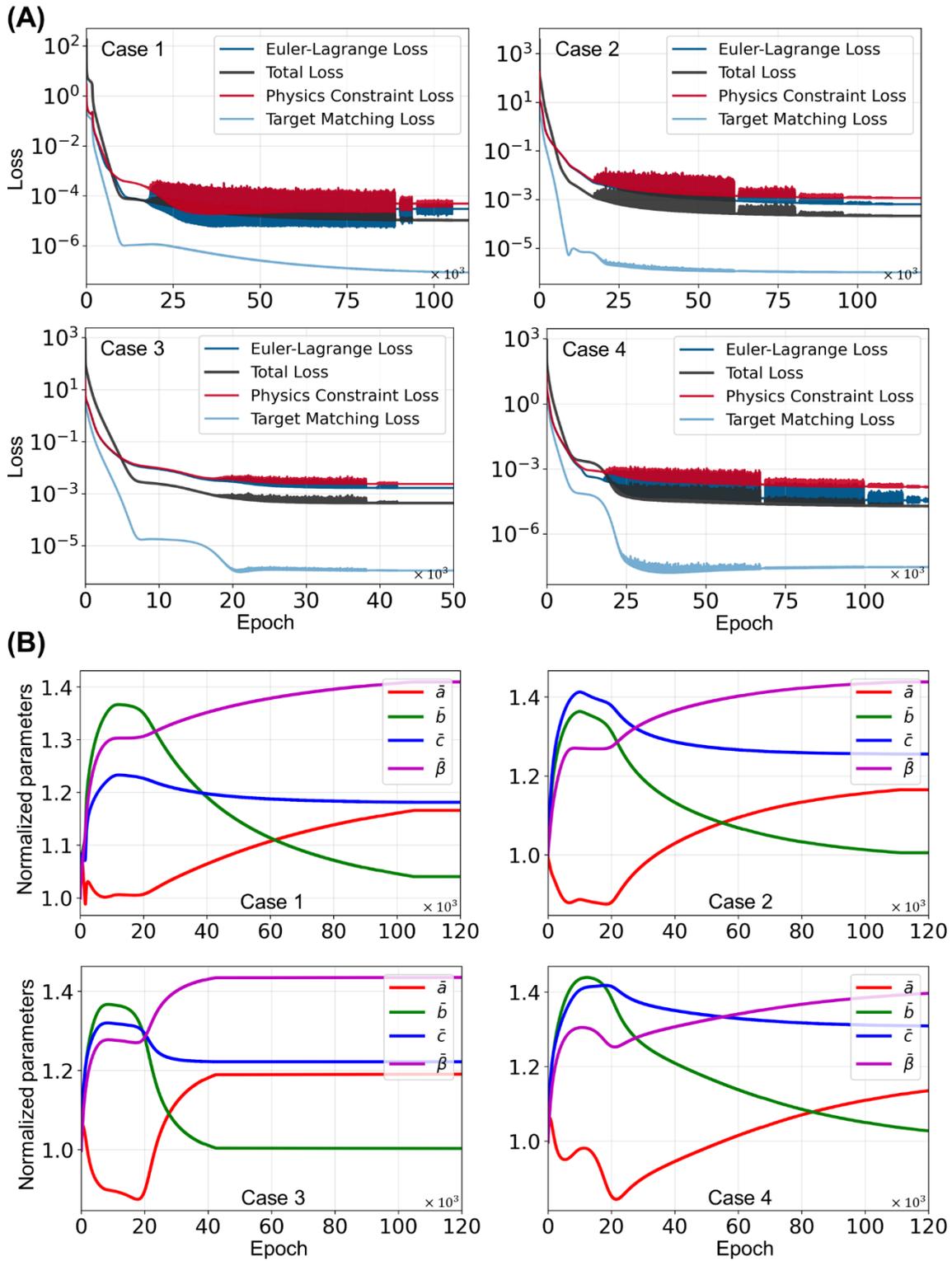

**Figure 6.** Convergence behavior during inverse design training. (A) Evolution of loss components over training epochs for the four representative cases from Figure 5. (B) Normalized structural parameters $(a, b, c, \beta)$ plotted against training epochs, showing stable convergence trends.

### 3.3. Inverse design based on bistable energy programming

We further assess the flexibility of the proposed inverse design framework by considering an underdetermined setting in which only minimal energy information is specified. In this bistable programming task, the input consists of two zero-energy stable heights and a target energy barrier, and the objective is to identify a physically feasible CKO geometry that satisfies these constraints.

To directly visualize the target stable positions and barrier height, we use the raw values of energy $U$ and height $h$ instead of normalized variables. Figure 7(A) shows a representative case with a design domain $h \in [0,4]$ mm, with stable points at $h_1 = 0$ mm and $h_2 = 3.9$ mm, and the target energy barrier set to 0.075 mJ. The inverse-designed energy curve (blue) matches the target profile, reproducing both zero-energy minima and the barrier height, and closely agrees with FEA validation (black dots) with $R^2 = 0.9987$ and $RMSE = 0.000964$.

Training convergence is illustrated in Figures 7(B) and (C). All structural parameters $(a, b, c, \beta)$ evolve smoothly toward stable values, and the total loss and its components decrease rapidly in the initial epochs and then gradually refine to reach stable values. These convergence plots confirm the numerical robustness of the inverse design framework for this challenging underdetermined problem. The comparison between the target and inverse-designed energy values at the stable points and barrier is summarized in Supplementary Table S3 (Supplementary Information Section 4), showing excellent agreement across all cases. The normalized absolute error remains below 1.4% in most configurations and does not exceed 2.3%, confirming the accuracy of the inverse design framework.

Figures 7(D) and (E) further demonstrate the flexibility and programmability of the framework. In Figure 7(D), five different cases are considered to evaluate performance under varying design targets. The first three cases fix the barrier height at 0.05 mJ and vary the second stable height: $h_2 = 3.4, 3.9$, and $4.4$ mm, with design domains [0,3.5], [0,4], and [0,4.5] mm, respectively. The remaining two fix $h_2 = 3.9$ mm and vary the barrier height: 0.05, 0.075, and 0.1 mJ. In all cases, the PINN-generated curves match the specified targets, and the corresponding 3D geometries at the barrier and second stable point are shown to the right of each plot.

In Figure 7(E), we test whether the same energy targets $(h_1 = 0\ mm, h_2 = 3.9\ mm, \text{barrier} = 0.075\ mJ)$ can be achieved while varying the number of polygon units in the CKO geometry: $n = 8, 9, 10$. The resulting energy curves again satisfy the targets, and FEA validation yields $R^2 > 0.99$ in all cases (Supplementary Table S5). Full parameter values for all inverse-designed cases are provided in Supplementary Table S4. Although the proposed framework demonstrates high flexibility, not all target barrier values can be physically realized. For a fixed pair of stable points $h_1$ and $h_2$, there exist upper and lower bounds dictated by the mechanics of the CKO unit, as discussed in Supplementary Information Section 5.

These results demonstrate that the PINN framework can perform inverse design of bistable energy profiles under minimal specifications, enabling precise and programmable control over both stable positions and barrier heights, even in highly constrained design spaces.

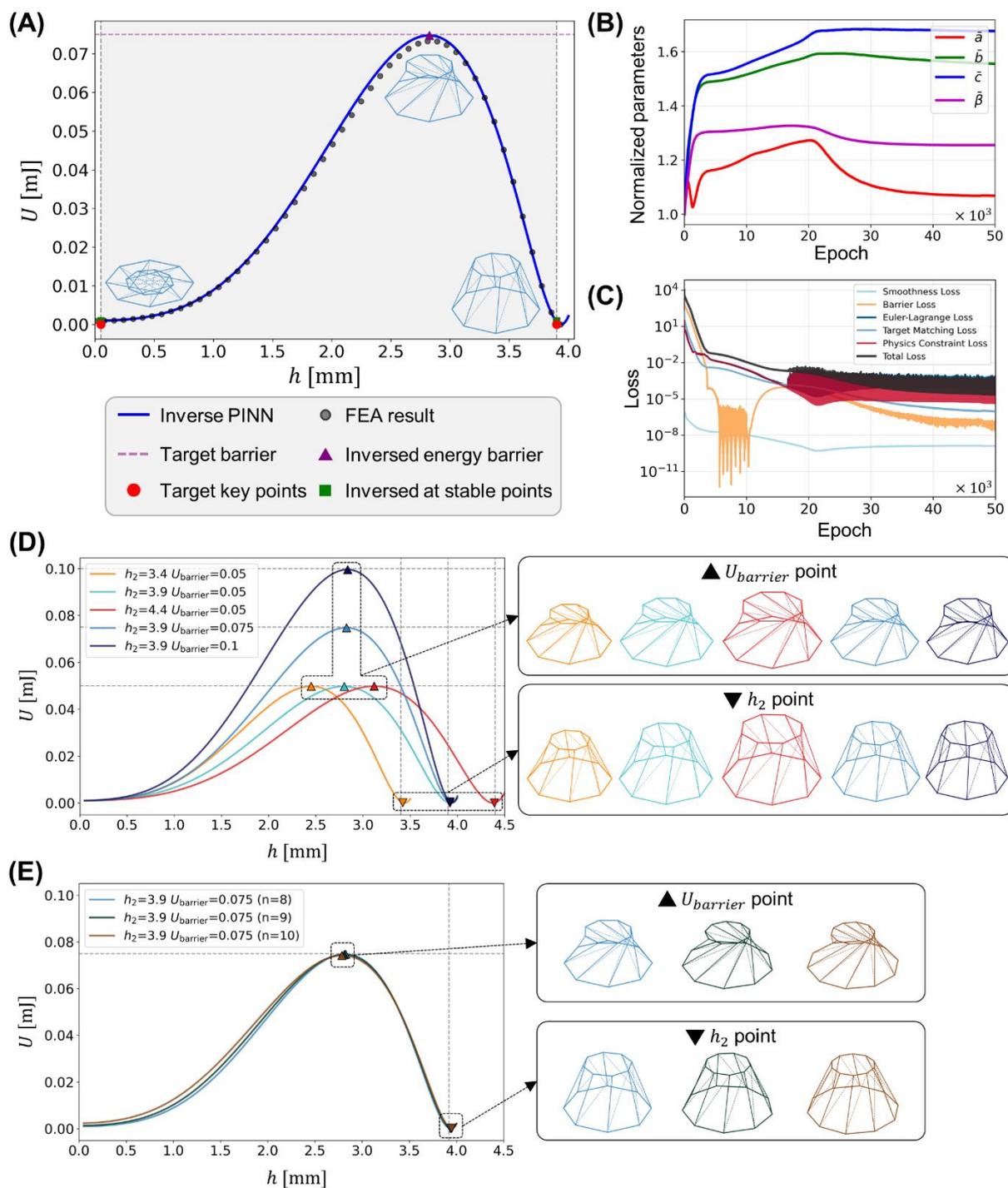

**Figure 7.** Bistable energy programming with minimal input. (A) Representative case: $h_1 = 0, h_2 = 3.9, U_{barrier} = 0.075$ mJ, domains [0,4]. (B) Evolution of normalized parameters $(a, b, c, \beta)$ for case in (A). (C) Loss convergence for case in (A). (D) Five design variations with $n = 8$: three cases with $h_2 = 3.4, 3.9, 4.4$ (domains [0,3.5], [0,4], [0,4.5]; $U_{barrier} = 0.05$ mJ) and two cases with $U_{barrier} = 0.075, 0.1$ mJ (domain [0,4]; $h_2 = 3.9$). Right panels show 3D structures at barrier and second stable states. (E) Polygon number variation ($n = 8, 9, 10$) for identical targets: $h_1 = 0, h_2 = 3.9, U_{barrier} = 0.075$ mJ, domain [0, 4]. Right panels show corresponding 3D structures.

### 3.4. Sequential deployment in multi-layer CKO structures via inverse design

We finally extend the inverse design framework to multi-layer CKO assemblies, aiming to achieve sequential deployment by prescribing distinct energy barriers for each layer. This capability enables complex shape changes without active control, as each layer folds in a predetermined order dictated by its barrier height.

To illustrate the concept, we design three configurations: a cylinder, a cone, and a zigzag, each composed of three vertically stacked CKO units. To ensure sequential bottom-up folding, the energy barrier heights of the three layers are set to 0.05 (bottom), 0.08 (middle), and 0.11 (top), respectively. Figure 8A shows the inverse-designed energy curves for all layers in each configuration, and the corresponding geometrical parameters are provided in Supplementary Table S6. For each shape, the top and bottom radii of the units are fixed to achieve the intended overall profile.

The compression-induced energy evolution for all three configurations is presented in Figure 8B. In every case, folding initiates in the bottom layer, proceeds to the middle, and concludes with the top layer, matching the design targets. This staged collapse pattern is reproduced in FEA simulations, with strong agreement between simulated and designed energy landscapes. Supplementary Movie 1 shows the dynamic deformation sequences. These results highlight that by shaping each layer's energy landscape and constraining its geometry, the framework can orchestrate multi-step deployments without active actuation. The same approach can be scaled to larger assemblies or adapted for temporal coordination in soft robotics and morphing structures.

To validate this multi-layer strategy, we fabricated and tested the zigzag configuration at a 16× scale based on the values in Supplementary Table S6, to facilitate practical fabrication and testing. Figure 9A shows the test setup for individual layers, alongside measured force–displacement and derived energy–displacement curves. As intended, the experimental barrier heights increase from Layer 1 to Layer 3, aligning with design targets (0.05, 0.08, 0.11 mJ). Quantitatively, the relative errors between normalized experimental values and PINN predictions are -7.93% for the middle layer and -16.16% for the top layer (Table 2). While the qualitative sequence is preserved, quantitative differences arise from physical effects absent in the truss-based model, such as crease rotation, polygon bending, and panel buckling, as observed in Supplementary Movie 2. These mechanisms dissipate additional energy, leading to residual energy after each folding event rather than a full return to zero as predicted by the model. The cumulative effect appears in both individual layer tests and the complete assembly.

Figure 9(B) presents the experimental validation of the complete three-layer zigzag assembly. The force–displacement curve reveals three distinct peaks corresponding to the sequential folding of each layer, with peak magnitudes increasing from bottom to top. The accompanying energy profile confirms the progressive accumulation effect observed in individual tests, wherein the energy does not return to zero between folding events but stabilizes at elevated levels. This behavior preserves the intended sequential deployment in the designed order (i.e., from the bottom to the top layer), with progressively increasing energy barriers, thereby validating our design strategy. Supplementary Movie 2 provides a dynamic visualization of the compression experiments, capturing the deformation of each individual layer and the complete multi-layer assembly. In the multi-layer test, the sequential folding sequence is clearly evident, with each layer folding in the predetermined order.

Overall, these experiments confirm that despite simplified modeling assumptions, the inverse-designed energy landscapes translate effectively to real structures, enabling robust sequential deployment with programmable barrier hierarchies.

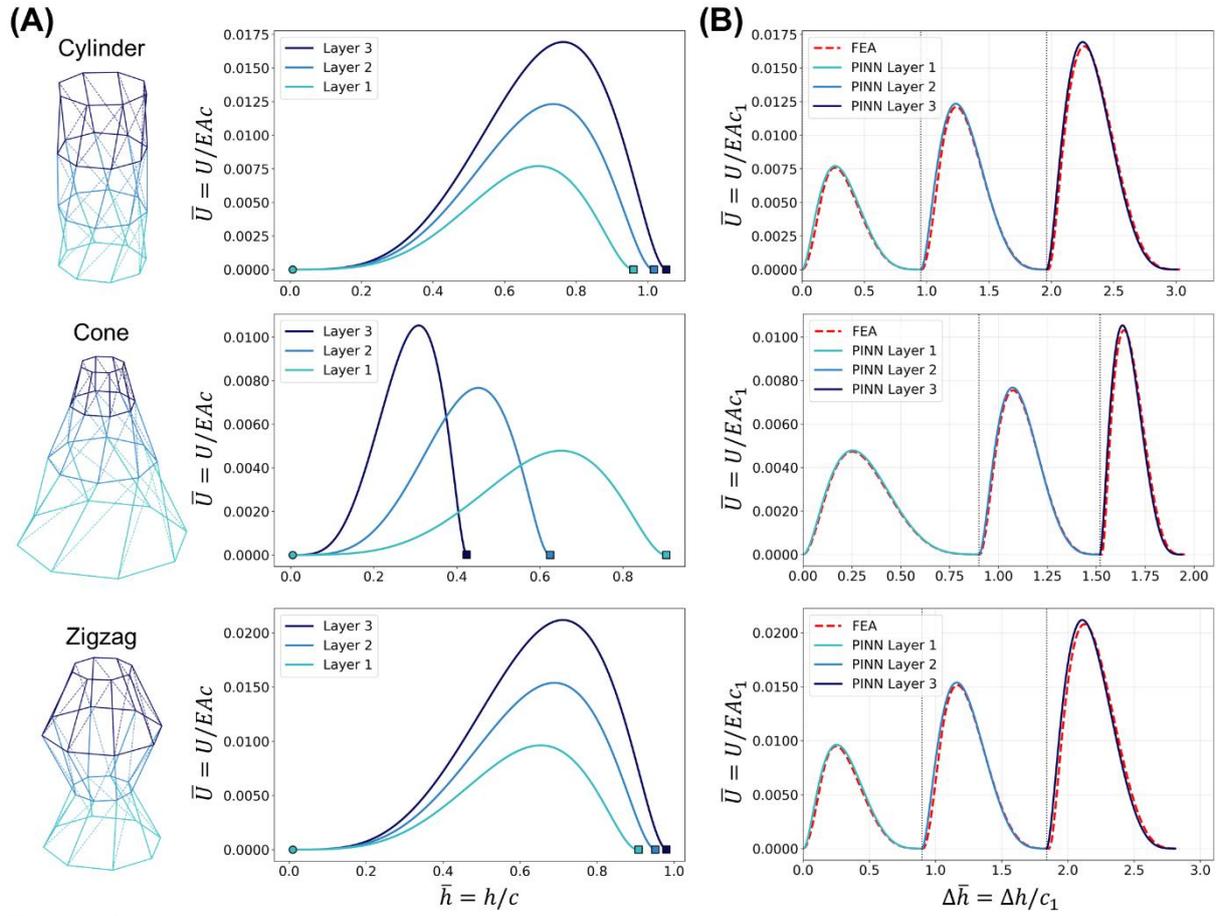

**Figure 8.** Multi-layer CKO structures with programmable sequential deployment. (A) Three target geometries (cylinder, cone, zigzag) showing 3D structures and inverse-designed energy curves $\bar{U}(\bar{h})$ for individual layers. Energy barriers: 0.05 (bottom), 0.08 (middle), 0.11 (top). (B) Energy evolution $\bar{U}(\Delta\bar{h})$ during compression. Solid lines: PINN; dashed lines: FEA.

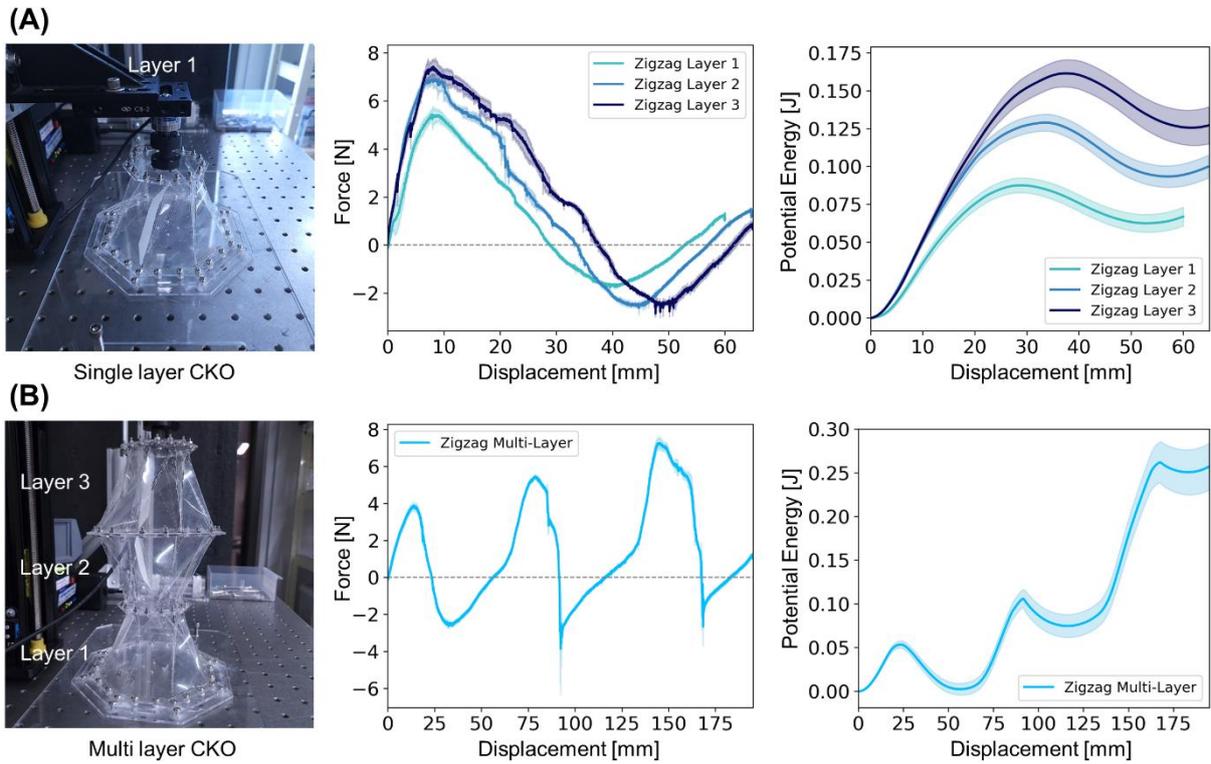

**Figure 9.** Experimental validation of zigzag configuration. (A) Individual layer tests: Layer 1 specimen photograph at initial state with force-displacement and energy-displacement curves for all three layers. (B) Three-layer assembly test: initial state photograph with force and energy response during compression showing three distinct peaks and energy accumulation.

| Layer | PINN prediction | Experimental [mJ] | Normalized experimental | Relative error (%) |
|---|---|---|---|---|
| Bottom | 0.05 | 0.0875 | 0.05 | - |
| Middle | 0.08 | 0.1289 | 0.0737 | -7.93 |
| Top | 0.11 | 0.1614 | 0.0922 | -16.16 |

**Table 2.** Comparison of PINN predictions and experimental energy barriers for each layer. Relative error is based on normalized experimental values; bottom layer is excluded as the normalization reference.

## 4. Discussion

This study establishes a comprehensive PINN framework for designing CKO with programmable mechanical behavior. The framework unifies forward prediction and inverse design without pre-collected training datasets, addresses longstanding challenges in modeling multistable origami mechanics, and clarifies how computational assumptions map to observed physical responses. Together with FEA and benchtop experiments, the results show that energy landscapes, such as stable states, barrier heights, and deployment paths, can be shaped directly through geometry.

The physics-informed strategy differs fundamentally from conventional data-driven approaches: instead of learning from labeled examples, it embeds mechanical equilibrium and energy principles into the loss so that predictions satisfy torque-free conditions and variational consistency by construction. These inductive biases improve sample efficiency and interpretability and enable extrapolation to regions of design space not explicitly seen during optimization. Extending the formulation from cylindrical to conical geometries further broadens the attainable design space by allowing different top/bottom radii and, consequently, richer kinematic couplings between axial motion and twist.

A key contribution is bistable energy programming from minimal specifications. Traditional inverse design often requires full target curves or many constraint points; our method needs only two stable heights and a desired barrier value, infers the barrier location automatically, and recovers the full energy landscape consistent with physics. This reduces the cognitive and computational burden on the designer while retaining control over deployment forces and states. Importantly, the same idea scales to hierarchical assemblies: by prescribing layer-wise barrier magnitudes, we achieve predictable, bottom-up folding in multi-layer CKO systems without active feedback or complex controls. From a practical standpoint, the results suggest a simple design rule: to ensure robust sequencing under real-world dissipation, maintain a sequence margin between adjacent barrier heights that exceeds expected experimental deviations (e.g., the top layer in our tests showed the largest underestimation). Such margins help preserve the intended order even when non-ideal effects accumulate.

Experiments highlight both strengths and limitations of the current model. The truss-based abstraction captures deployment order and relative barrier magnitudes, but quantitative differences arise because real prototypes exhibit deformation modes omitted in the model: rotational stiffness at creases, bending of polygonal frames, panel buckling, frictional/contact effects, and associated dissipation. These mechanisms explain the observed energy accumulation between folding events, whereas the idealized model predicts a return to zero. A natural next step is to enrich the energy functional with hinge torsion and frame/panel bending and twisting terms, while retaining computational tractability. Two complementary routes are attractive: (i) a gray-box extension that augments the analytic energy with a low-dimensional learned discrepancy term regularized by physics, and (ii) multi-fidelity training, in which the PINN is calibrated to FEA but nudged toward experiment with sparse measurements. Both approaches can reduce bias while preserving the interpretability and stability benefits of physics-informed learning.

Beyond model form, there are methodological considerations. Inverse design is one-to-many: multiple geometries can realize the same energy curve. Our framework currently selects one feasible solution; in practice, designers may prefer solutions that optimize secondary criteria such as manufacturability (feature sizes, angle ranges), robustness to tolerances, or material economy. These can be incorporated as soft penalties or additional objectives (e.g., minimum-norm parameters, sensitivity-aware regularization that penalizes large barrier gradients with respect to $a, b, c, \beta, n$). Likewise, initialization and nonconvexity can affect convergence; multi-start strategies, curriculum schedules for loss weights, and parameter nondimensionalization are pragmatic safeguards that we found to promote stable optimization. Finally, while our study focuses on quasi-static behavior, many applications involve rate effects and dynamics; extending the framework to time-dependent PINNs with damping and inertia, or to hybrid simulators that combine our energy model with reduced-order dynamics, would enable trajectory-level design (e.g., speed of deployment, impact mitigation).

The broader implications extend to mechanical metamaterials and deployable structures in which energy landscape design is the primary lever. Aerospace systems can benefit from passive, fail-safe sequencing in booms, reflectors, and morphing skins; soft robotics can exploit programmed barriers to realize snap-through actuation, mechanical memory, or force-limited behaviors; impact-mitigation devices can tune threshold energies for staged absorption. Because the formulation is fast, differentiable, and interpretable, it lends itself to design-automation workflows and interactive tools that allow rapid exploration of alternatives and real-time prediction. Looking forward, coupling the framework with uncertainty quantification (e.g., Bayesian PINNs or ensemble surrogates) would enable reliability-aware designs that maintain sequencing under fabrication variability. Incorporating fabrication-aware constraints (minimum crease length, perforation patterns, allowable thickness) and contact/friction models will further close the gap to practice.

## 5. Conclusions

This work presents a PINN framework that addresses both forward prediction and inverse design of CKO without relying on pre-collected training data. By embedding mechanical equilibrium and energy principles directly into the learning objective, the framework produces interpretable, physically consistent designs across a broad range of geometries and operating conditions.

First, we develop a data-free learning approach that predicts and designs complex energy landscapes governed by multistability. In forward prediction, the framework accurately captures monostable and bistable behaviors across diverse geometries, closely matching FEA with $R^2 \geq 0.93$ and small RMSE values on the order of $10^{-3}$. In inverse design against full target curves, agreement is even tighter, with $R^2 \geq 0.99$ relative to both target profiles and FEA validations and RMSE below $10^{-3}$. Together, these results demonstrate high fidelity while eliminating the need for labeled datasets.

Second, we introduce an inverse-design methodology that programs energy barriers from minimal specifications, namely two stable heights and a target barrier value, while autonomously inferring the barrier location and recovering a full, physically feasible energy profile. The method generalizes across stable-state spacing, barrier magnitude, and polygon count $n$, and identifies feasible solutions even in underdetermined settings. We also delineate that realizable barrier heights are bounded for a fixed pair of stable states, clarifying practical design limits.

Third, we extend the approach to hierarchical, multi-layer assemblies and realize programmable sequential deployment by prescribing layer-wise barrier magnitudes. FEA confirms bottom-to-top folding in the intended order for cylinder, cone, and zigzag configurations. Benchtop experiments on a three-layer zigzag prototype (scaled 16×) reproduce the designed sequencing and relative barrier hierarchy. The relative errors of the normalized experimental barriers are -7.93% and -16.16% for the middle and top layers, respectively. While quantitative differences arise from deformation modes not included in the truss model (crease torsion, frame bending, panel buckling, friction/contact), the qualitative deployment order and barrier progression are preserved, supporting real-world robustness.

The established framework provides a foundation for fabrication-aware, physics-consistent design of origami-inspired metamaterials. Immediate directions include enriching the energy functional with hinge torsion and frame/panel bending, incorporating contact/friction models, and adopting gray-box or multi-fidelity training to reduce model–experiment bias while retaining interpretability. Integrating uncertainty quantification and sensitivity-aware penalties can improve robustness to tolerances and preserve sequencing margins under variability. Extending to time-dependent settings (damping/inertia) will enable trajectory-level design for rate-sensitive applications.

In summary, the proposed PINN framework turns geometric parameters into programmable energy landscapes, ranging from single units to hierarchical assemblies, and bridges theory and practice in origami engineering. Its efficiency, generality, and experimental viability point to broad impact across deployable aerospace structures, soft-robotic actuators, impact-mitigation devices, and biomedical mechanisms where precise, adaptable mechanical responses are essential.

**Author Contributions**

Sukheon Kang: Conceptualization, Data curation, Formal analysis, Investigation, Methodology, Software, Validation, Visualization, Writing - original draft, Writing – review & editing.

Youngkwon Kim: Formal analysis, Investigation, Methodology, Validation, Writing – review & editing.

Jinkyu Yang: Formal analysis, Methodology, Writing – review & editing.

Seunghwa Ryu: Conceptualization, Funding acquisition, Methodology, Project administration, Supervision, Writing - review & editing.

**Data availability**

Data will be made available on request

**Conflicts of interest**

The authors declare that they have no known competing financial interests or personal relationships that could have appeared to influence the work reported in this paper.

**Acknowledgements**

This work was supported by the National Research Foundation of Korea (NRF) (No. RS-2023-00222166) and the InnoCORE program (No. N10250154), both funded by the Ministry of Science and ICT (MSIT), as well as by a grant from the Ministry of Food and Drug Safety (No. RS-2023-00215667). Y.K. and J.Y. acknowledge the support from the Air Force Office of Scientific Research (FA2386-24-1-4051) and the National Research Foundation of Korea (2023R1A2C2003705 and No. 2022H1D3A2A03096579).

# Supplementary information

# Physics-informed neural networks for programmable origami metamaterials with controlled deployment

Sukheon Kang [a], Youngkwon Kim [b], Jinkyu Yang [b,*], and Seunghwa Ryu [a,c,*]

**Affiliations**

[a] Department of Mechanical Engineering, Korea Advanced Institute of Science and Technology, 291 Daehak-ro, Yuseong-gu, Daejeon 34141, Republic of Korea

[b] Department of Mechanical Engineering, Seoul National University, 1 Gwanak-ro, Gwanak-gu, Seoul 08826, Republic of Korea

[c] KAIST InnoCORE PRISM-AI Center, Korea Advanced Institute of Science and Technology, 291 Daehak-ro, Yuseong-gu, Daejeon 34141, Republic of Korea

*Corresponding author e-mail: ryush@kaist.ac.kr (S. Ryu)

*Corresponding author e-mail: jkyang11@snu.ac.kr (J. Yang)

**Section 1. Geometric and energetic modeling of conical Kresling origami**

This section provides a detailed description of the geometric definitions, coordinate expressions, and energy formulation for the conical Kresling origami (CKO) structure. These formulations support the truss-based modeling introduced in Section 2.1 of the main manuscript and form the foundation for the physics-informed learning framework.

The CKO structure consists of $n$ identical unit cells arranged in a periodic manner. Each cell is defined by five geometric parameters: the top and bottom edge lengths $a$ and $b$, the mountain and valley crease lengths $c$ and $d$, and the internal angle $\beta$ between the bottom edge and its adjacent mountain crease. The folded structure forms a conical shell in which the top and bottom polygons are inscribed in circles of radius $r$ and $R$, respectively, given by:

$$r = \frac{a}{2\sin(\frac{\pi}{n})} \tag{S1}$$

$$R = \frac{b}{2\sin(\frac{\pi}{n})} \tag{S2}$$

where $n$ is the number of unit cells. The folding configuration is described by two state variables: the vertical height $h$ between the polygons and the relative rotational angle $\varphi$ between them. The spatial coordinates of the vertices are then defined in cylindrical coordinates as follows, where $i = 1, 2, \ldots, n$

$$A_i = \begin{bmatrix} r\cos(\frac{2\pi(i-1)}{n} + \varphi) \\ r\sin(\frac{2\pi(i-1)}{n} + \varphi) \\ h \end{bmatrix} \tag{S3}$$

$$B_i = \begin{bmatrix} R\cos(\frac{2\pi(i-1)}{n}) \\ R\sin(\frac{2\pi(i-1)}{n}) \\ 0 \end{bmatrix} \quad (S4)$$

Using these coordinates, the deformed lengths of the mountain and valley creases, denoted $\tilde{c}(h,\varphi)$ and $\tilde{d}(h,\varphi)$, respectively, can be computed as the Euclidean distances between corresponding vertices, where $c$ and $d$ represent the initial lengths of the mountain and valley creases.

$$\tilde{c}(h,\varphi) = |A_i - B_i| = \sqrt{h^2 + r^2 + R^2 - 2rR\cos(\varphi)} \quad (S5)$$

$$\tilde{d}(h,\varphi) = |A_{i+1} - B_i| = \sqrt{h^2 + r^2 + R^2 - 2rR\cos(\varphi + \frac{2\pi}{n})} \quad (S6)$$

These expressions describe the nonlinear coupling between the vertical displacement and rotational deformation of the CKO structure.

Using the deformed crease lengths, the total strain energy of the CKO structure can be expressed as the sum of the elastic strain energies stored in all mountain and valley creases:

$$U(h,\varphi) = \sum_{i=1}^{n} [\frac{nk_m}{2}(\tilde{c}(h,\varphi) - c)^2 + \frac{nk_v}{2}(\tilde{d}(h,\varphi) - d)^2] \quad (S7)$$

where $k_m = \frac{EA}{c}$ and $k_v = \frac{EA}{d}$ denote the axial stiffness of the mountain and valley creases, respectively. Here, $E$ is the Young's modulus and $A$ is the cross-sectional area of the crease elements. This energy expression provides a foundation for evaluating equilibrium states, stiffness, and energy barriers between configurations.

The equilibrium states of the structure for a given vertical height $h$ correspond to torque-free configurations, obtained by enforcing:

$$\frac{\partial U}{\partial \varphi} = 0 \tag{S8}$$

Solving Eq. (S8) yields the equilibrium rotation profile $\varphi(h)$, which can then be substituted back into Eq. (S7) to compute the minimum-energy path. This minimum-energy trajectory corresponds to the red dashed curves shown in Figure 1(D) of the main text and forms the basis for the energy–height profiles illustrated in Figure 1(E).

These formulations fully describe the geometric and energetic behavior of the CKO structure within the truss-based modeling framework. They provide the physically consistent, data-free foundation for the physics-informed neural network (PINN) approach used in the forward prediction and inverse design tasks described in the main manuscript.

## Section 2. Algorithms and Implementation Details

This section provides detailed algorithms and implementation specifications for the physics-informed neural network (PINN) framework used in forward prediction and inverse design of conical Kresling origami structures. Three algorithms are presented: Algorithm 1 for forward prediction, Algorithm 2 for inverse design based on target energy curves, and Algorithm 3 for inverse design with bistable energy programming.

### 2.1 Network Architecture and Implementation

We employed different network architectures for forward prediction and inverse design problems. For forward prediction (Algorithm 1), we utilized a fully connected neural network consisting of two hidden layers with 128 neurons each, activated by hyperbolic tangent functions. The input layer receives only the sampled height vector $\boldsymbol{h}$, while the geometric parameters $(a, b, c, \beta, n)$ are treated as fixed constants inside the energy model. The output layer employs a sigmoid-based transformation to enforce physical constraints on the rotation angle $\varphi$, mapping it to the admissible range $[\varphi_{lower}, \varphi_{upper}]$, where $\varphi_{upper} = min(\varphi_f, \varphi_s)$ represents the minimum of the flat-folding and locked configuration angles.

For inverse design problems (Algorithms 2 and 3), the framework comprises two coupled models: a PhysicsNN for predicting the rotation field and a ParameterModel for optimizing structural parameters. The PhysicsNN maintains the same architecture as the forward prediction network with two hidden layers of 128 neurons each. The ParameterModel employs parameter-specific transformations to ensure physical validity: logarithmic transformations for positive parameters a and b, and sigmoid transformations for constrained parameters $c \in [|r - R|, r + R]$ and $\beta \in (0, \pi)$.

## 2.2 Loss Function Components and Their Roles

The loss functions in our framework consist of multiple components, each serving specific purposes:

1. Energy Minimization Loss ($L_E$): Drives the network to minimize the total potential energy, ensuring physical accuracy.

2. Euler-Lagrange Residual Loss ($L_{EL}$): Promotes smoothness in the predicted rotation profile $\varphi(h)$ and prevents abrupt curvature changes or nonphysical jumps in the resulting energy curve.

3. Physics Equilibrium Loss ($L_{phys}$): Minimizes rotational torque ($\partial U/\partial \varphi = 0$) and ensures static equilibrium across the deformation path.

4. Target Matching Loss ($L_{target}$): Enforces consistency between the predicted energy curve and the given target profile.

5. Barrier Constraint Loss ($L_{barrier}$): Ensures that the energy gradient vanishes at the inferred barrier point ($L_{barrier,1}$) and that the barrier is a unique maximum through convexity checking ($L_{barrier,2}$).

6. Smoothness Loss ($L_{smooth}$): Regularizes the energy curve globally, preventing sharp curvature transitions.

7. Constraint Penalty ($L_{penalty}$): A hard constraint enforcement mechanism that applies a large penalty ($C = 10^3$) when geometric or physical feasibility constraints are violated.

## 2.3 Training Hyperparameters

The training hyperparameters were carefully selected based on the complexity of each task. For forward prediction (Algorithm 1), we employed the Adam optimizer with a learning rate of 1e-3. The loss function incorporates an Euler-Lagrange regularization term with weight $\lambda_{EL} = 0.1$ and a smoothness coefficient $\lambda_{reg} = 0.01$.

For inverse design based on full energy curves (Algorithm 2), we used a learning rate scheduler (ReduceLROnPlateau) with a reduction factor of 0.5 and patience of 100 epochs. The multi-objective loss function was weighted as follows: $\lambda_{target} = 20.0$ for target energy matching, $\lambda_{phys} = 0.1$ for physics equilibrium constraints, and $\lambda_{EL} = 0.1$ for variational consistency.

For bistable energy programming (Algorithm 3), additional loss terms were introduced to control the energy landscape. The weights were set as $\lambda_{target} = 10.0$, $\lambda_{phys} = 1.0$, $\lambda_{EL} = 0.1$, $\lambda_{smooth} = 0.05$ for global smoothness, and $\lambda_{barrier} = 5.0$ for barrier energy constraints.

All computations were performed using single-precision floating-point arithmetic (float32) to balance computational efficiency with numerical accuracy. Physical constants were normalized with E = A = 1 for dimensional consistency. Constraint violations were penalized with a large constant C = 1000 to effectively restrict the parameter space to physically realizable configurations.

## 2.4 Algorithms

---

**Algorithm 1** Forward design of origami structure using PINN

---

**Input:** Structural parameters $(a, b, c, \beta, n)$, domain $[h_{min}, h_{max}]$

**Output:** Energy curve $U(h)$, rotation angle $\varphi(h)$

1: Initialize PINN model to predict $\varphi(h)$

2: Compute physical constraints: $\varphi_{lower} = 0$, $\varphi_{upper} = \min(\varphi_f, \varphi_s)$

3: **repeat**

4:   Predict $\varphi(h)$ using PINN

5:   Compute energy $U(h, \varphi(h))$ using truss model

6:   Compute physical loss: $\partial U/\partial \varphi = 0$

7:   Compute 1st derivative: $\partial \varphi/\partial h$

8:   Compute 2nd derivative: $\partial^2 \varphi/\partial h^2$

9:   Compute EL residual: $\partial U/\partial \varphi - 2\lambda_{reg}(\partial^2 \varphi/\partial h^2)$

10:   Compute EL loss: $L_{EL} = \text{mean}[(\partial U/\partial \varphi - 2\lambda_{reg}(\partial^2 \varphi/\partial h^2))^2]$

11:   Compute total loss: $L_{total} = L_E + \lambda_{EL} L_{EL}$

12:   Update PINN parameters using gradient descent

13: **until** Max epoch is reached;

14: Return $U(h, \varphi(h))$, $\varphi(h)$

---

**Algorithm 2** Inverse design of origami structure based on target energy curve using PINN

---

**Input:** Initial structural parameters $(a, b, c, \beta, n)$, Target energy curve $U_{target}(h)$

**Output:** Optimized structural parameters $(a, b, c, \beta)$

1: Initialize parameter model and physics model (PINN)
2: **repeat**
3:     Get current structural parameter $(a, b, c, \beta)$ from parameters model
4:     Check physical validity of parameters
5:     Predict $\varphi(h)$ using physics model
6:     Compute energy $U(h, \varphi(h))$ using truss model
7:     Compute target matching loss: $L_{target} = \text{mean}[(U(h) - U_{target}(h))^2]$
8:     Compute physical loss: $L_{phys} = \text{mean}[(\partial U/\partial \varphi)^2]$
9:     Compute EL loss as in Algorithm 1
10:    Compute total loss: $L_{total} = \lambda_{target} L_{target} + \lambda_{phys} L_{phys} + \lambda_{EL} L_{EL} + L_{penalty}$
11:    Update PINN parameters using gradient descent
12: **until** Max epoch is reached;
13: Return optimized structural parameters $(a, b, c, \beta)$

**Algorithm 3** Inverse design of origami structure with bistable energy programming using PINN

---

**Input:** Initial structural parameters $(a, b, c, \beta, n)$, Target energy barrier value $U_{barrier}$,
Two stable points $(h_1, U_1 = 0), (h_2, U_2 = 0)$

**Output:** Optimized structural parameters $(a, b, c, \beta)$

1: Initialize parameter model and physics model (PINN)
2: Find indices of stable points $h_1$ and $h_2$ in discretized domain
3: Define barrier search region: $[min(h_1, h_2), max(h_1, h_2)]$
4: **repeat**
5:     Get current structural parameters $(a, b, c, \beta)$ from parameter model
6:     Check physical validity of parameters
7:     Predict $\varphi(h)$ using physics model
8:     Compute energy $U(h, \varphi(h))$ using truss model
9:     Extract energy values at stable points
10:     Compute stable points matching loss: $L_{target} = \text{mean}[(U(h_i) - U_{target}(h_i))^2]$ for i ∈ {1,2}
11:     Compute physical loss: $L_{phys} = \text{mean}[(\partial U/\partial \varphi)^2]$
12:     Compute EL loss as in Algorithm 1
13:     Compute smoothness loss: $L_{smooth} = \text{mean}[(\partial^2 U/\partial \varphi^2)^2]$
14:     Find energy maximum in barrier search region: $h_{max} = argmax_{h \in [h_1, h_2]} U(h)$
15:     Compute barrier gradient constraint: $L_{barrier,1} = \left|\frac{\partial U}{\partial h}\right|^2_{h=h_{max}}$
16:     Compute barrier convexity constraint: $L_{barrier,2} = \text{ReLU}(\frac{\partial^2 U}{\partial h^2})_{h=h_{max}}$
17:     Compute total barrier loss: $L_{barrier} = L_{barrier,1} + L_{barrier,2}$
18:     Compute total loss: $L_{total} = \lambda_{target} L_{target} + \lambda_{phys} L_{phys} + \lambda_{EL} L_{EL} + \lambda_{smooth} L_{smooth}$
                              $+\lambda_{barrier} L_{barrier} + L_{penalty}$
19:     Update PINN and parameter model using gradient descent
20: **until** Max epoch is reached;
21: Return optimized structural parameters $(a, b, c, \beta)$

**Section 3. Inverse design based on full energy curves - Detailed parameters**

This section provides detailed structural parameters and accuracy metrics for the inverse design framework based on full energy curves (Section 3.2 of the main manuscript). The parameters correspond to the four representative cases shown in Figure 5.

| Case | Parameter [mm] | Target | Inverse PINN |
|---|---|---|---|
| Case 1 | $a$ | 1 | 1.3210 |
| | $b$ | 2 | 1.6450 |
| | $c$ | 3.1 | 2.9653 |
| | $\beta$ [rad] | 1.5 | 1.6542 |
| Case 2 | $a$ | 1 | 1.3977 |
| | $b$ | 2 | 1.5078 |
| | $c$ | 3.5 | 3.1377 |
| | $\beta$ [rad] | 1.5 | 1.7253 |
| Case 3 | $a$ | 1 | 1.3773 |
| | $b$ | 2 | 1.5339 |
| | $c$ | 3.3 | 3.0579 |
| | $\beta$ [rad] | 1.5 | 1.6962 |
| Case 4 | $a$ | 1 | 1.3611 |
| | $b$ | 2 | 1.5427 |
| | $c$ | 3.5 | 3.2718 |
| | $\beta$ [rad] | 1.5 | 1.6748 |

**Supplementary Table S1.** Structural parameters for inverse design of CKO geometries. Target parameters used to generate reference energy curves and inverse-designed parameters obtained through the PINN framework for cases in Figure 5.

| Case | $R^2$ (Target) | RMSE (Target) | $R^2$ (FEA) | RMSE (FEA) |
| --- | --- | --- | --- | --- |
| Case 1 | 0.993539 | 0.000097 | 0.998506 | 0.000045 |
| Case 2 | 0.999589 | 0.000321 | 0.991991 | 0.001347 |
| Case 3 | 0.999947 | 0.000039 | 0.997255 | 0.000270 |
| Case 4 | 0.999925 | 0.000047 | 0.998268 | 0.000217 |

**Supplementary Table S2.** Accuracy metrics for inverse design based on full energy curves. $R^2$ and RMSE values comparing PINN predictions with target energy curves and FEA validation results for the four cases in Figure 5.

**Section 4. Additional results for inverse design with bistable energy programming**

This section presents comprehensive validation results for the inverse design framework with bistable energy programming across seven different design scenarios. We systematically varied the second stable point height ($h_2$), target energy barrier values ($U_{barrier}$), and the number of polygon units ($n$) to demonstrate the robustness and versatility of our approach. All figures and tables in this section are arranged in the following order: (1) $h_2$ = 3.4 mm, $U_{barrier}$ = 0.05 mJ; (2) $h_2$ = 3.9 mm, $U_{barrier}$ = 0.05 mJ; (3) $h_2$ = 4.4 mm, $U_{barrier}$ = 0.05 mJ; (4) $h_2$ = 3.9 mm, $U_{barrier}$ = 0.075 mJ; (5) $h_2$ = 3.9 mm, $U_{barrier}$ = 0.1 mJ; (6) $h_2$ = 3.9 mm, $U_{barrier}$ = 0.075 mJ with n = 9; (7) $h_2$ = 3.9 mm, $U_{barrier}$ = 0.075 mJ with n = 10. Unless otherwise specified, n = 8 polygon units were used.

Supplementary Figure 1 shows the inverse-designed energy curves for all seven cases, each validated against finite element analysis (FEA) simulations. The first three cases ($h_2$ = 3.4, 3.9, and 4.4 mm) maintain a constant barrier height of 0.05 mJ while varying the second stable point location, demonstrating the framework's ability to control deployment distance. The next two cases explore different barrier heights (0.075 and 0.1 mJ) at a fixed $h_2$ = 3.9 mm, validating precise control over deployment forces. The final two cases examine the effect of polygon number (n = 9 and 10) while maintaining identical energy specifications, confirming the framework's adaptability to different geometric configurations.

The training convergence behavior is illustrated in Supplementary Figure 2, which displays the evolution of total loss and its components across all cases. All designs achieved stable convergence within 50,000 epochs, with the total loss typically reaching values below $10^{-4}$. The loss curves exhibit characteristic two-phase behavior: rapid initial descent followed by gradual refinement, indicating robust optimization dynamics. Notably, cases with higher polygon numbers (n = 9 and 10) demonstrated slower loss reduction compared to the n = 8 cases, suggesting that the increased geometric complexity requires more careful navigation of the optimization landscape to achieve the desired energy specifications.

Supplementary Figure 3 presents the evolution of structural parameters ($a, b, c, \beta$) during training for each case. The parameter trajectories demonstrate smooth convergence without oscillations, validating the effectiveness of our parameter transformation strategies. Interestingly, designs with different polygon numbers (n = 8, 9, 10) converged to distinct parameter sets while achieving identical energy specifications, highlighting the non-unique nature of the inverse design problem and the framework's ability to find valid solutions within the feasible design space.

Quantitative validation results are summarized in Supplementary Table 1. The agreement between PINN predictions and FEA simulations is excellent across all cases, with coefficients of determination ($R^2$) exceeding 0.998 and root mean square errors (RMSE) below 0.0016. The highest accuracy was achieved for the $h_2$ = 4.4 mm case ($R^2$ = 0.9993, RMSE = 0.000469), while the largest RMSE occurred for the highest barrier case ($U_{barrier}$ = 0.1 mJ), which can be attributed to the increased nonlinearity in the energy landscape. These metrics confirm that the physics-informed learning successfully captures the complex mechanical behavior across diverse design specifications.

The consistency of results across varying geometric configurations (n = 8, 9, 10) with identical energy targets demonstrates that our framework can accommodate different structural realizations of the same mechanical behavior. This flexibility is particularly valuable for practical applications where manufacturing constraints or integration requirements may dictate specific geometric configurations. Furthermore, the successful inverse design across a wide range of barrier heights (0.05 to 0.1 mJ) validates the framework's applicability to diverse deployment force requirements, from delicate biomedical devices to robust aerospace structures.

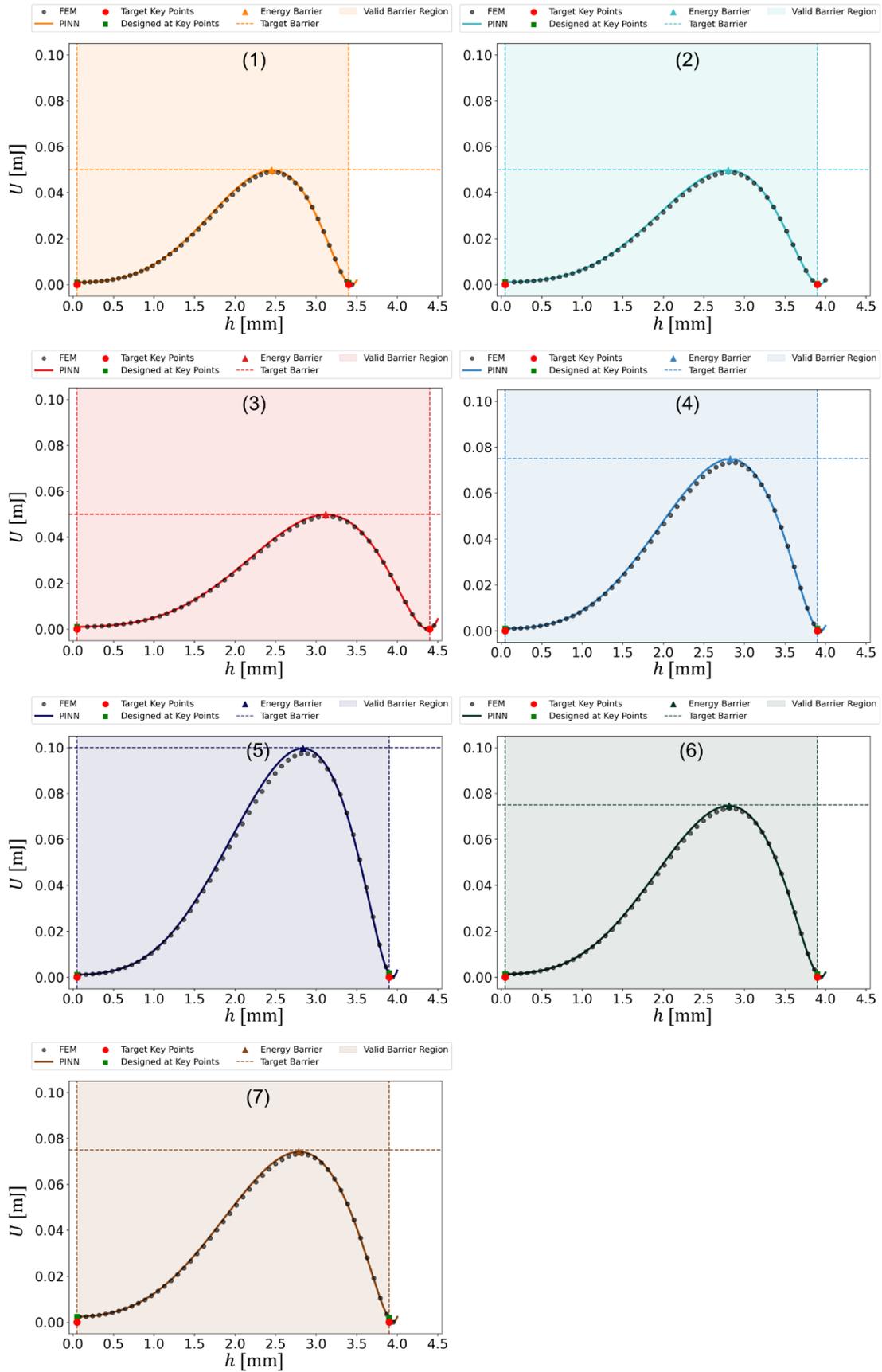

**Supplementary Figure S1.** Comparison of inverse-designed energy curves (lines) with FEA validation (dots) for seven bistable design cases.

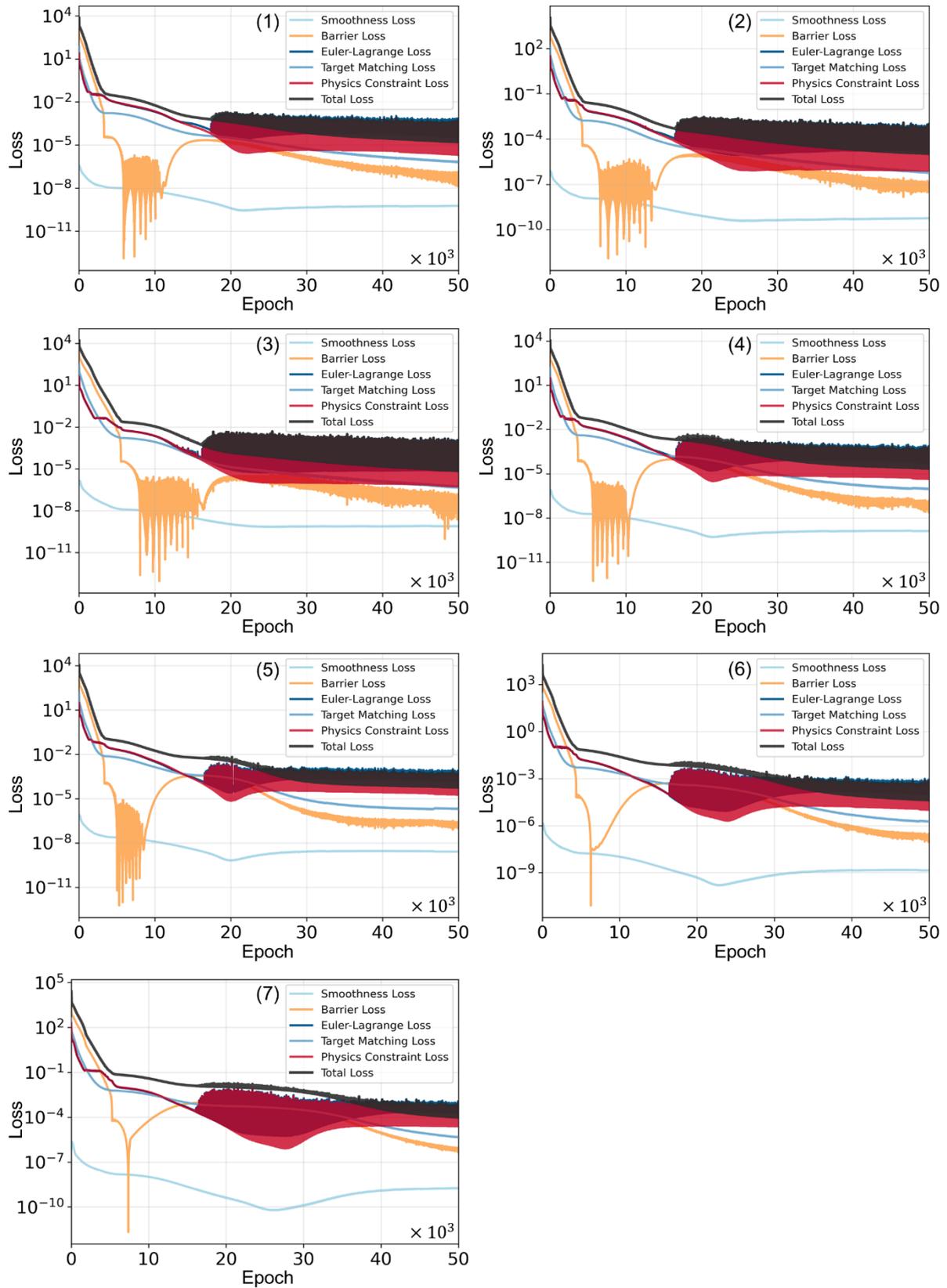

**Supplementary Figure S2.** Training loss evolution showing convergence of all loss components over 50,000 epochs for the seven inverse design cases.

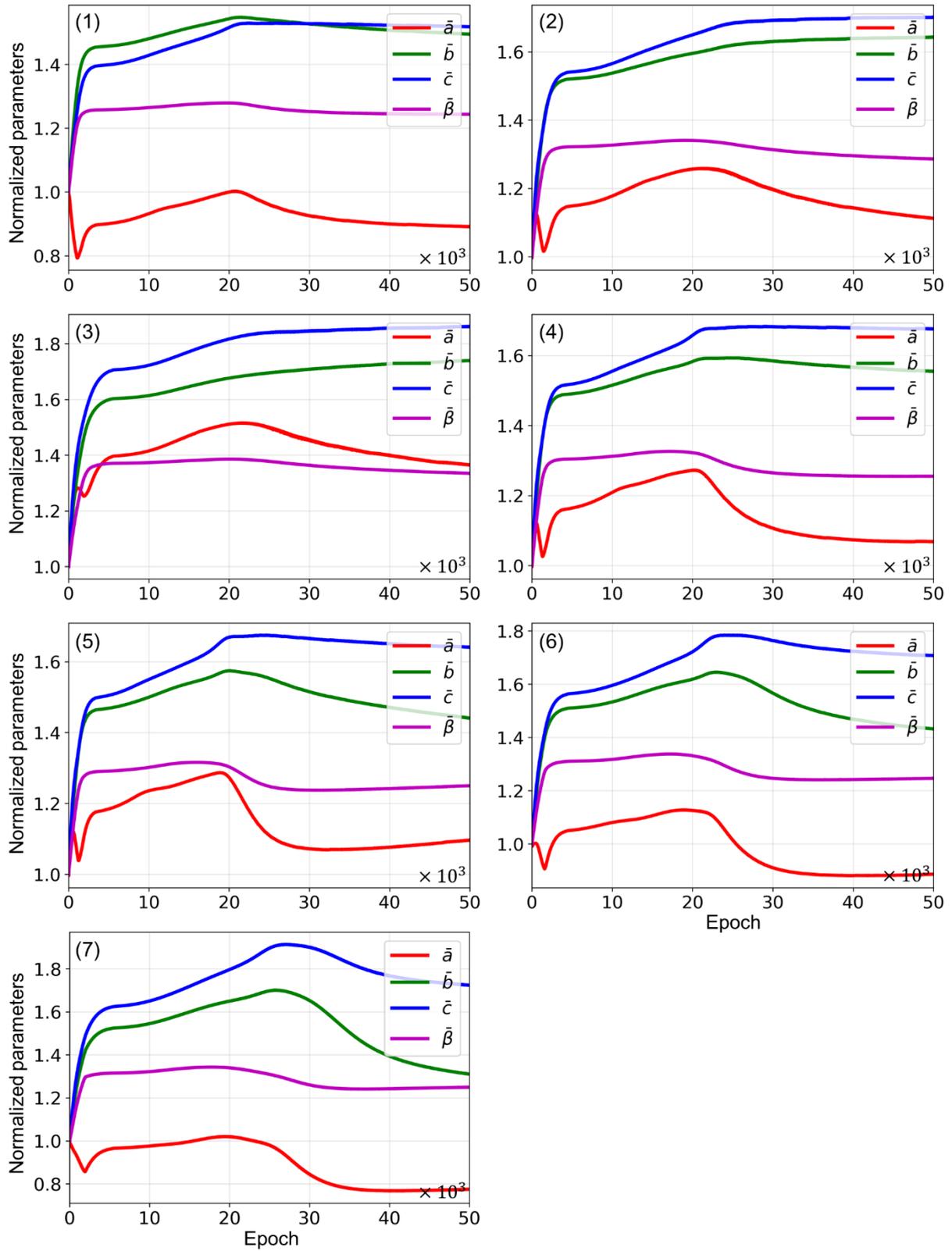

**Supplementary Figure S3.** Evolution of normalized structural parameters during inverse design training for the seven cases.

| Design targets | $h_1$ [mm] | $h_2$ [mm] | $U_{barrier}$ [mJ] | NAE (%) |
|---|---|---|---|---|
| $h_2 = 3.4$, $U_{barrier} = 0.05$, n = 8 | 0.000905 | 0.000691 | 0.049766 | 1.22 |
| $h_2 = 3.9$, $U_{barrier} = 0.05$, n = 8 | 0.001071 | 0.000243 | 0.049728 | 1.06 |
| $h_2 = 4.4$, $U_{barrier} = 0.05$, n = 8 | 0.000982 | 0.000089 | 0.049843 | 0.82 |
| $h_2 = 3.9$, $U_{barrier} = 0.075$, n = 8 | 0.000950 | 0.000959 | 0.074743 | 0.96 |
| $h_2 = 3.9$, $U_{barrier} = 0.1$, n = 8 | 0.001050 | 0.001725 | 0.099600 | 1.06 |
| $h_2 = 3.9$, $U_{barrier} = 0.075$, n = 9 | 0.001308 | 0.001394 | 0.074578 | 1.39 |
| $h_2 = 3.9$, $U_{barrier} = 0.075$, n = 10 | 0.002352 | 0.001887 | 0.074184 | 2.25 |

**Supplementary Table S3.** Target specifications and achieved values for bistable energy programming. Comparison of target and inverse-designed values for stable point heights ($h_1, h_2$), barrier energy ($U_{barrier}$), and normalized absolute error (NAE) for all cases in Figure 7.

| Design targets | Parameter [mm] | Value |
| --- | --- | --- |
| $h_2 = 3.4$, $U_{barrier} = 0.05$, n = 8 | a | 1.0693 |
| | b | 2.2421 |
| | c | 3.7955 |
| | $\beta$ [rad] | 1.4923 |
| $h_2 = 3.9$, $U_{barrier} = 0.05$, n = 8 | a | 1.3351 |
| | b | 2.4642 |
| | c | 4.2524 |
| | $\beta$ [rad] | 1.5434 |
| $h_2 = 4.4$, $U_{barrier} = 0.05$, n = 8 | a | 1.6380 |
| | b | 2.6089 |
| | c | 4.6543 |
| | $\beta$ [rad] | 1.6016 |
| $h_2 = 3.9$, $U_{barrier} = 0.075$, n = 8 | a | 1.2817 |
| | b | 2.3321 |
| | c | 4.1901 |
| | $\beta$ [rad] | 1.5061 |
| $h_2 = 3.9$, $U_{barrier} = 0.1$, n = 8 | a | 1.3156 |
| | b | 2.1610 |
| | c | 4.1020 |
| | $\beta$ [rad] | 1.4997 |
| $h_2 = 3.9$, $U_{barrier} = 0.075$, n = 9 | a | 1.0632 |
| | b | 2.1481 |
| | c | 4.2691 |
| | $\beta$ [rad] | 1.4955 |
| $h_2 = 3.9$, $U_{barrier} = 0.075$, n = 10 | a | 0.9292 |
| | b | 1.9656 |
| | c | 4.3098 |
| | $\beta$ [rad] | 1.4991 |

**Supplementary Table S4.** Structural parameters for bistable energy programming. Inverse-designed parameters for CKO structures with specified stable points and energy barriers, including variations in $h_2$, barrier height, and polygon number (n = 8, 9, 10) shown in Figure 7(D) and (E).

|  | $R^2$ (FEA) | RMSE (FEA) |
|---|---|---|
| $h_2 = 3.4$, $U_{barrier} = 0.05$, n = 8 | 0.9991 | 0.000531 |
| $h_2 = 3.9$, $U_{barrier} = 0.05$, n = 8 | 0.9990 | 0.000562 |
| $h_2 = 4.4$, $U_{barrier} = 0.05$, n = 8 | 0.9993 | 0.000469 |
| $h_2 = 3.9$, $U_{barrier} = 0.075$, n = 8 | 0.9987 | 0.000964 |
| $h_2 = 3.9$, $U_{barrier} = 0.1$, n = 8 | 0.9982 | 0.001506 |
| $h_2 = 3.9$, $U_{barrier} = 0.075$, n = 9 | 0.9991 | 0.000808 |
| $h_2 = 3.9$, $U_{barrier} = 0.075$, n = 10 | 0.9992 | 0.000706 |

**Supplementary Table S5.** Quantitative comparison between PINN predictions and FEA simulations for all seven inverse design cases.

**Section 5. Physical bounds on realizable energy barriers**

The inverse design framework with bistable energy programming, while highly flexible, cannot achieve arbitrary combinations of stable point locations and energy barrier values due to fundamental physical constraints. This section analyzes the bounds on achievable energy barriers for given stable point configurations.

For a bistable structure with stable points at $h_1$ and $h_2$, the realizable energy barrier has both lower and upper bounds. The lower bound represents the minimum energy required for any transition between stable states, arising from inevitable crease deformation. The upper bound is more restrictive and depends on geometric constraints and kinematic limits of the structure. Supplementary Figure 4 illustrates the characteristic failure modes when target specifications exceed physical limits. Two distinct behaviors emerge when attempting to design beyond the upper bound. In the first mode (Supplementary Figure 4(A)), the optimization successfully maintains zero energy at stable points $h_1$ and $h_2$ but fails to achieve the target barrier height, resulting in a plateau below the desired value. In the second mode (Supplementary Figure 4(B)), the framework achieves the target barrier height but violates the stability constraints, producing non-zero energy at the intended stable points. The training dynamics reveal these limitations clearly. As shown in the loss evolution plots, the optimization initially converges smoothly, but upon approaching physical limits, the loss suddenly diverges. This divergence indicates that the required structural parameters would violate geometric constraints, such as c exceeding r+R or becoming less than |r-R|, or would demand kinematically impossible configurations.

These bounds have direct implications for design practice. Target specifications must be verified against physical limits before optimization. If convergence fails with behavior similar to Supplementary Figure 4(A), the stable point separation should be increased to enable higher barriers. If failure resembles Supplementary Figure 4(B), the barrier requirement is fundamentally incompatible with the specified stable points and must be reduced. Understanding these limits prevents wasted computational effort and guides designers toward feasible specifications.

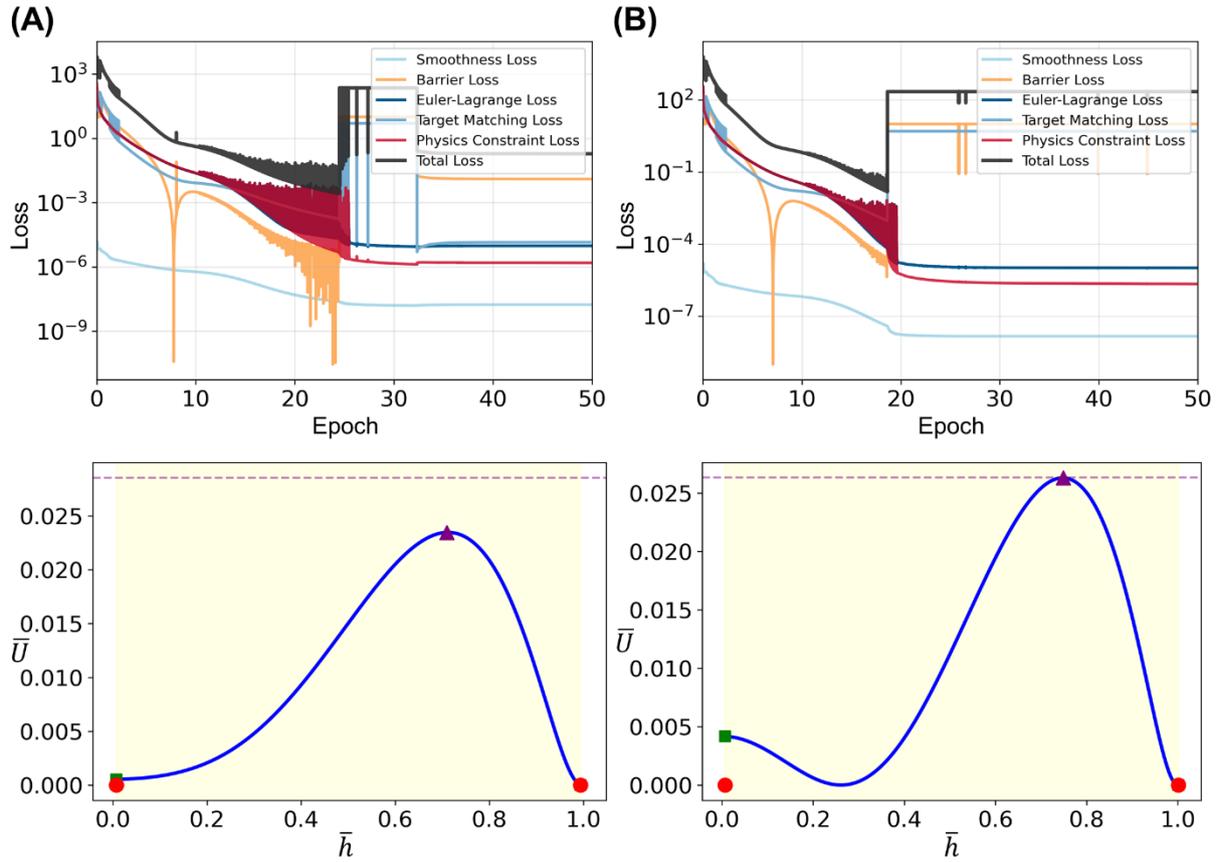

**Supplementary Figure S4.** Inverse design results demonstrating physical limitations. (A) Loss evolution and energy profile for a case where the target barrier height cannot be achieved while maintaining zero energy at stable points. (B) Loss evolution and energy profile for a case where zero energy at stable points cannot be maintained while achieving the target barrier height.

## Section 6. Multi-layer CKO assemblies - Design parameters

This section provides the detailed design parameters for the multi-layer CKO assemblies with programmable sequential deployment presented in Section 3.4 of the main manuscript.

| Design targets | Parameter [mm] | Value |
|---|---|---|
| Cylinder layer 1 (U=0.05) | a | 3.0615 |
| | b | 3.0615 |
| | c | 6.4995 |
| | $\beta$ [rad] | 1.7983 |
| Cylinder layer 2 (U=0.08) | a | 3.0615 |
| | b | 3.0615 |
| | c | 6.7476 |
| | $\beta$ [rad] | 1.7433 |
| Cylinder layer 3 (U=0.11) | a | 3.0615 |
| | b | 3.0615 |
| | c | 6.9247 |
| | $\beta$ [rad] | 1.7009 |
| Cone layer 1 (U=0.05) | a | 3.8268 |
| | b | 6.1229 |
| | c | 10.4319 |
| | $\beta$ [rad] | 1.6587 |
| Cone layer 2 | a | 2.2961 |
| (U=0.08) | b | 3.8268 |
| | c | 6.9116 |
| | $\beta$ [rad] | 1.5736 |
| Cone layer 3 (U=0.11) | a | 1.5307 |
| | b | 2.2961 |
| | c | 4.5472 |
| | $\beta$ [rad] | 1.5219 |
| Zigzag layer 1 (U=0.05) | a | 1.5307 |
| | b | 3.0615 |
| | c | 5.1951 |
| | $\beta$ [rad] | 1.5294 |
| Zigzag layer 2 (U=0.08) | a | 1.5307 |
| | b | 3.0615 |
| | c | 5.3689 |
| | $\beta$ [rad] | 1.4924 |
| Zigzag layer 3 (U=0.11) | a | 1.5307 |
| | b | 3.0615 |
| | c | 5.4883 |
| | $\beta$ [rad] | 1.4626 |

**Supplementary Table S6.** Design parameters for multi-layer CKO assemblies with programmed sequential deployment. Structural parameters for three-layer configurations (cylinder, cone, zigzag) with hierarchical energy barriers of 0.05, 0.08, and 0.11 for bottom, middle, and top layers, respectively.